\documentstyle[11pt,aaspp4]{article}

\slugcomment{Submitted to ApJ}

\lefthead{Hannikainen et al.}
\righthead{GRO J1655--40 Radio Observations}

\def\groj1655{{\rm GRO\,J1655$-$40}}

\begin{document}

\title{Radio Emission from GRO\,J1655--40 during the 1994 Jet Ejection Episodes}

\author{D. C. Hannikainen}
\affil{Observatory, PO Box 14, FIN-00014 University of Helsinki,
         Finland}

\author{R. W. Hunstead and D. Campbell-Wilson}
\affil{School of Physics, University of Sydney, NSW 2006, Australia}

\author{K. Wu}
\affil{Special Research Centre for Theoretical Astrophysics, School of Physics,
             University of Sydney, NSW 2006, Australia}

\author{D. J. McKay\altaffilmark{1,2}}
\affil{University of Manchester, Nuffield Radio Astronomy Laboratories,
         Jodrell Bank, \\
Cheshire SK11~9DL, United Kingdom}

\author{D. P. Smits}
\affil{Department of Mathematics, Applied Mathematics \& Astronomy,
University of South Africa, PO~Box 392, Pretoria 0003, South Africa}

\and

\author{R. J. Sault}
\affil{Australia Telescope National Facility, CSIRO, PO Box 76, Epping,
              NSW 1710, Australia}

\altaffiltext{1}{Joint Institute for VLBI in Europe, Postbus 2, 
           7990 AA Dwingeloo, The Netherlands}
\altaffiltext{2}{Australia Telescope National Facility, 
  CSIRO --- Paul Wild Observatory, 
  Locked Bag 194,
  Narrabri, NSW 2390,
  Australia}

\begin{abstract}

We report multifrequency radio observations of \groj1655 obtained with
the Australia Telescope Compact Array, the Molonglo Observatory
Synthesis Telescope and the Hartebeesthoek Radio Astronomy Observatory
at the time of the major hard X-ray and radio outbursts in
1994 August-September.  The radio emission reached levels of the order
of a few Jy and was found to be linearly polarized by up to 10\%,
indicating a synchrotron origin.  The light curves are in good
agreement with those measured with the VLA, but our closer time
sampling has revealed two new short-lived events and significant
deviations from a simple exponential decay.  The polarization data
show that the magnetic field is well ordered and aligned at
right angles to the radio jets for most of the monitoring period.  The
time evolution of the polarization cannot be explained solely in
terms of a simple synchrotron bubble model, and we invoke a
hybrid `core-lobe' model with a core which contributes both
synchrotron and free-free emission and `lobes' which are classical 
synchrotron emitters.

\end{abstract}


\section{Introduction}\label{intro}

The X-ray transient \groj1655 (Nova Sco 1994) was first detected with
the {\em Burst and Transient Source Experiment} (BATSE) on board the
Compton Gamma-Ray Observatory on 1994 July 27 (\cite{nan94}).
Significant flux was observed up to 200 keV and the source reached a
luminosity of 1.1 Crab (20--100 keV) on Aug 1. \groj1655 remained in
outburst until about Aug 15, and after a period of quiescence flared
again on Sept 6 (\cite{harmon95}).  The detection of a steeply-rising
radio counterpart was first reported by Campbell-Wilson \& Hunstead
(1994a) following observations with the Molonglo Observatory Synthesis
Telescope (MOST) at 843~MHz on 1994 Aug 6 and 11.  The flux density
continued to increase, reaching 4.2 Jy on 1994 Aug 14 and 5.5 Jy on
1994 Aug 15 (\cite{cw&h94b}).  High resolution radio observations
using the Very Large Array (VLA), the Very Long Baseline Array (VLBA)
(Hjellming \& Rupen 1995, hereafter HR95) and the Southern Hemisphere
VLBI Experiment (SHEVE) array (\cite{tin95}) showed repeated episodes
of relativistic ejections. Three major ejection events on TJD~9577.5,
9584 and 9596 (TJD defined as JD $-$ 2440000.5) were observed with the 
VLA and three on TJD~9574$\pm$1,
9605$\pm$2 and 9668$\pm$5 were recorded with the VLBA (\cite{h&r95}).
The ejection velocity, after correction for inclination to the line of
sight, was inferred to be $v\sim0.92c$ in the rest frame of \groj1655,
at a position angle of $47^{\circ}\pm1^{\circ}$ (\cite{h&r95});
wiggles about the jet axis were interpreted as precession with a
period of $3.0\pm0.2$d.  Photometric observations showed \groj1655
to be a high-inclination binary system (\cite{bailyn95}). Subsequent 
optical observations led to the determination of a precise mass
for the primary, $M_1=7.02\pm0.22\,M_{\odot}$ (\cite{o&b97}), which is
well above the theoretical upper limit for a neutron star and direct
evidence for it being a black hole. In addition, Soria et al.\ (1998)
found a 95\% confidence limit of $M_1\,>\,5.1\,M_{\odot}$ for the mass of
the primary based on measurements of velocity variations in
the He~{\sc ii} disk emission lines, thought to reflect the orbital
motion of the primary, and hence confirming \groj1655 to be a
black hole binary. The mass of the secondary, which is classified as
{\rm F}3 {\rm IV}--{\rm F}6 {\rm IV}, is $M_2=2.34\pm0.12\,M_{\odot}$,
and the spectroscopic period of the binary is
$P=2^d.62157\pm0^d.00015$ (\cite{o&b97}).
 
During the 1994 ejection events, the radio outbursts from \groj1655
were monitored in the southern hemisphere by the Molonglo Observatory
Synthesis Telescope (MOST), the Australia Telescope Compact Array
(ATCA), and the Hartebeesthoek Radio Astronomy Observatory (HartRAO).
We discuss below the results of these observations.  Sections
\ref{obs+dr} and \ref{lightcurves} describe the MOST, ATCA and HartRAO
observations and the resulting light curves.  In Section \ref{linpol}
we present the linear polarization data from the ATCA, and we discuss
the evolution of the radio spectra in Section \ref{spectralstuff}.  We
briefly compare our results with those of HR95 in Section
\ref{compare}.  Finally, we interpret the polarization data in terms
of a `core-lobe' model in Section \ref{physint}.

\section{Observations and data reduction}\label{obs+dr}

\subsection{MOST}

The MOST is an east-west synthesis array comprising two colinear
cylindrical paraboloids each 11.6~m wide by 778~m long, separated by
a 15~m gap (\cite{mills81}; \cite{rob91}).  The telescope operates at
843~MHz, with a detection bandwidth of 3.25~MHz, and measures right
circular polarization (as defined by the IEEE standard). The
synthesized beamwidth is $43\arcsec\times43\arcsec {\rm cosec} |\delta|$
FWHM (RA $\times$ Dec).  The telescope forms a comb of 64 real-time
fan beams spaced by $22\arcsec$; for the observations of \groj1655 the
pointing of the beam set was time shared among three adjoining
positions to give a field size of $70\arcmin\times70\arcmin$
cosec$|\delta|$.  The background noise in a full 12-hour synthesis
image was typically 1.5 mJy rms.

Many \groj1655 observations were partial syntheses as the observations
were fitted in around a previously scheduled observing program. The
choice of field center serendipitously included a strong point source
($\approx$ 900 mJy, J2000 position $16^{\rm h} 53^{\rm m} 55.9^{\rm
s}, -40^{\arcdeg} 37^{\arcmin} 23^{\arcsec}$) that has subsequently
been used as an internal reference source.  The flux density
calibration was based on short fan-beam {\sc scan} observations of
strong point sources from the list of Campbell-Wilson and Hunstead
(1994c).  For S$_{\rm 843\,MHz}>100$~mJy the dominant source of error
in a synthesis image is the scatter in the pre- and post-observation
calibrators; a conservative error of $\pm 3$\% has been assigned. At
lower flux densities the error becomes dominated by noise and, for
partial syntheses, residuals arising from incomplete cancellation of
strong out-of-field sources.  This difficulty can be partly overcome
by subtracting a reference image (obtained when \groj1655 was
quiescent) with the same hour angle coverage.

The data processing used standard MOST reduction software.  When the
source was bright we were able to search for sample-to-sample (24~s)
variations in flux density.  Such measurements are affected by
confusing sources in MOST's fan beams, by the effects of ionospheric
refraction and by slow drifts in telescope calibration.  These effects
are eliminated or greatly reduced in a full 12-hour synthesis.  For
this paper, individual sample flux densities were fitted and
averaged in blocks of 200 (80 minutes) and then combined in 4--6-hour
blocks for tabulation.  Full synthesis imaging was used when the
source flux density fell below 500~mJy, with flux densities measured
using IMFIT in {\sc aips}.

In addition to the synthesis-mode observations of \groj1655, spot
measurements of flux density were obtained using 4-minute {\sc scan}
observations, bracketed by {\sc scan}s of calibrators.  This mode of
operation measures the target and calibrator source flux densities
using the (same) central beams of the 64-beam block. It was used only
when the flux density of \groj1655 was greater than 1~Jy, with the fit
parameters serving as a guide to the presence of confusing sources in
the fan-beam response.  The final light curve, tabulated in
Table~\ref{tbl-most}, contains a mix of flux densities measured from
synthesis images, from averaged fits to the synthesis data, and from
{\sc scan} observations, and includes data from a third, weaker
outburst which occurred in 1994 Nov (\cite{w&h97}; \cite{hann}).

\placetable{tbl-most}

\subsection{ATCA}

Following the report of the intense radio outburst by the MOST,
\groj1655 was monitored as a target-of-opportunity with the 
ATCA for a period of about twenty days.
VLBI observations and \ion{H}{1} spectral line observations were also
undertaken (\cite{m&k94}; \cite{tin95}).

The ATCA is an earth-rotation aperture synthesis array, comprising six
22-m antennas which can be moved along an east-west railway track to
give baselines up to 6~km (\cite{frater}). 
The observations of \groj1655 were made at central frequencies of
1.380, 2.378, 4.800, 5.900, 8.640 and 9.200~GHz with 128~MHz bandwidth
in two orthogonal linear polarizations. Each observation was typically
10 minutes duration, with the antenna gain and phase calibration
derived from regular observations of the point-source calibrator
PKS~B1740$-$517.  The flux density scale was tied to the primary
calibrator PKS~B1934$-$638, which we assume has flux densities of
14.96, 11.54, 5.83 and 2.84~Jy at 20, 13, 6 and 3~cm (1.4, 2.3, 4.8,
8.6~GHz) respectively (\cite{rey94}).  The 6.0A configuration was
used throughout, giving interferometer spacings from 337--5939~m.

The data were reduced using standard flagging and calibration
techniques in the {\sc miriad} package (\cite{bob95}). The
polarimetric calibration of ATCA data is described in \cite{bob91}.
Polarimetric leakages, which are quite stable with time on the ATCA,
were determined simultaneously with the antenna gains using either
PKS~B1934$-$638 or PKS~B1740$-$517, both known to be unpolarized.
The resultant polarimetric purity of the calibrated data is believed
to be better than 0.1\%. At all wavelengths, a single round of phase
self-calibration was performed on the \groj1655 data (using a point
source model) to eliminate residual phase instability. As the
observations placed \groj1655 at the phase center, we simply summed
the real parts of the relevant Stokes visibilities to determine the
source flux density; this is equivalent to natural weighting.  
The errors are dominated by systematic effects. 
For the total intensity 
measurements, the  errors are smaller than 1\% of Stokes~I,
while for Stokes~Q and U they are less than 0.1\% of Stokes~I.
The ATCA total flux densities are tabulated in Table~\ref{tbl-atca}. 

\placetable{tbl-atca}

\subsection{HartRAO}

Continuum measurements of the flux density of \groj1655 were made at
3.5 and 6.0~cm using the 26-m telescope of the Hartebeesthoek Radio
Astronomy Observatory (HartRAO) situated in Gauteng, South Africa.
The observations were made using dual-feed systems operated in a
beam-switched Dicke mode.  At 3.5~cm (8.58 GHz) the bandwidth was 400
MHz, the beamwidth $5.7'$ FWHM and the feed recorded right circular
polarization.  The 6~cm (5 GHz) receiver had a bandwidth of 800 MHz, a
beamwidth of $10'$ FWHM and the feed was linearly polarized east-west.

The flux density at each wavelength was measured using a five-point
stepping sequence in declination to correct for pointing errors in
this coordinate; the stepping positions in terms of beam parameters
were first null, half-power, on-source, half-power, first null. The
five points were fitted to a Gaussian to determine declination
pointing errors.  Right ascension errors were measured at 6~cm by
observing midway between the beams. Right ascension pointing
corrections at 3.5~cm were assumed to be the same as at 6~cm. The beam
separation was sufficiently close ($15'$ or 1.5 times the
beamwidth at 6~cm) that the response of the dual-beam Dicke system
varied linearly with right ascension in this region.  The observed
flux density is an average over both beams, corrected for both
declination and right ascension pointing errors. Measurements
alternated between the two wavelengths with 10 minutes' integration on
each.

Gain curve corrections, determined by G.\ Nicolson (private
communication), were applied to the data.  Flux densities were
calibrated to an arbitrary scale using a noise diode, and the
calibrators 3C\,123, 3C\,161, 3C\,218 (Hydra A), 3C\,274 (Virgo A) and
3C\,348 were observed occasionally to put the flux densities on the
Ott et al. (1994) scale.  The HartRAO data are tabulated in
Table~\ref{tbl-rao}; we note that on some days the scatter among
a closely spaced sequence of measurements is somewhat larger than
implied by their formal errors.

\placetable{tbl-rao}

\section{Radio light curves}\label{lightcurves}

\subsection{MOST and HartRAO}\label{most+rao_lc}

\placefigure{fig-most+batse}

The MOST 843~MHz light curve up to the end of 1994 September is
plotted in the top panel of Figure~\ref{fig-most+batse}.  The duration
of the ATCA observations is also shown, as are the epochs of
specific ejection episodes, as reported by \cite{h&r95} from their VLA
and VLBA data; we labelled these epochs {\sc vla1}, {\sc vla2}, {\sc
vla3} and {\sc vlba1}, {\sc vlba2}, corresponding to TJD 9577.5, 9584,
9596 and TJD~9574, 9605, respectively.  The BATSE 20--100 keV light
curve (provided by S.\ N.\ Zhang and W.\ S.\ Paciesas) is plotted in
the bottom panel, showing two major hard X-ray outbursts, beginning on
TJD~$\sim 9560$ and TJD~$\sim 9600$.

Two radio outbursts were observed with the MOST in 1994
August--September, both apparently associated with the hard X-ray
bursts detected by BATSE. The MOST data clearly show the rise of the
radio flux density lagging behind the rise of the X-ray intensity in
the outbursts. During the first X-ray outburst, the radio flux density
was still low (S$_{\rm 843\,MHz}$ $\simeq$ 0.36 Jy) on TJD~9570.60,
while the X-ray intensity had already reached its maximum. When the
X-ray intensity began to decline, the radio flux density then
increased rapidly, reaching a local maximum (S$_1$) of 5.45 Jy on
TJD~9579.22.  After a brief decline, the flux density rose again and
reached a new peak of 7.62 Jy on TJD~9582.13, about two weeks after
the X-ray peak.  At this point the X-ray intensity had already fallen
to a much lower level, but was still above the background. 
The radio peak in the second outburst (S$_2$) also lagged the X-ray
peak, but the delay was shorter.  The maximum intensities in the two
X-ray bursts were similar, but the second radio burst was obviously
weaker.

\subsubsection{Short Duration Radio Burst Component}\label{short} 

\placefigure{fig-most+rao}

To emphasize some specific features of the temporal evolution of the
radio emission, we plot in Figure~\ref{fig-most+rao} (upper panel) the
MOST 843~MHz and HartRAO 5~GHz flux densities on a logarithmic scale
as a function of time.  Closely spaced groups of 5~GHz points are
plotted as weighted averages.  The 843~MHz flux density curve is
characterized by short-duration burst components (marked S$_1$ and
S$_2$) superimposed on longer-lived components (L$_1$ and L$_2$) which
show an approximately exponential decay.  HartRAO did not observe
\groj1655 until after the end of S$_1$, but S$_2$ was clearly recorded
at 5~GHz.  The confirmation of the short duration burst component is
an important new result, since neither the VLA (\cite{h&r95}) nor the
ATCA collected data during these periods.  From the MOST data, the
short duration components, S$_1$ and S$_2$, have e-folding rise times
of $\leq1.1$ days, and S$_2$ decays with an e-folding time of $\sim
2.2$ days.  The rising time scale for the long duration component
cannot be determined from the data, but L$_1$ and L$_2$ have e-folding
decay times of $\sim 6.5$ days and $\sim 8.7$ days respectively.  A
third short duration event was observed by MOST peaking on TJD~$\sim
9666$ which decayed with an e-folding time of $\sim 2.6$ days
(Table~\ref{tbl-most}; \cite{w&h97}).  In addition, the flare observed
in 1996 had an initial e-folding decay time of $\sim 1.4$ days,
similar in scale to the times quoted above (\cite{hunt97}).  It is
tempting to speculate that event S$_2$ may be a prompt response to the
brief hard X-ray event at TJD 9610.  A similar association at TJD 9579
can be made for S$_1$ but seems less plausible.

\subsection{ATCA}\label{atca_lc}

\placefigure{fig-atca}

Figure~\ref{fig-atca} shows the ATCA Stokes I light curves at all six
observing frequencies, again plotted on a log-linear scale to
emphasize the overall exponential decay; the MOST light curve is
included for comparison.

The ATCA observations were initiated during the decay phase of the S1
short duration burst component.  It is clear from
Figure~\ref{fig-atca} that the flux density at all six ATCA
frequencies declines simultaneously with the MOST flux density between
TJD~$\sim 9579-9580$ suggesting that the short burst component was
present at {\it all\/} frequencies.  Following the minimum at
TJD~$\sim 9580$ the flux density increases for all six ATCA
frequencies, reaching a local maximum of 5.49~Jy at 1.4~GHz on
TJD~9581.62; note, however, that the overall maximum flux density at
frequencies $\geq4.8$~GHz occurs {\it prior\/} to the minimum.  For
TJD~$>9582$ there is a steady decline, with time constants similar to
MOST's at the lower frequencies but shorter at the higher frequencies
(4.8--9.2 GHz).  Other features, such as the dip at the higher
frequencies around TJD~9583 and the rise between 9596 and 9598 at
8.6~GHz, are discussed in later sections.

\section{Linear polarization}\label{linpol} 

Linear polarization was detected at all six ATCA frequencies
throughout the epoch of the observations. The fractional polarization
at each frequency, ${\rm P/I}\,=\,\sqrt{({\rm Q}^2+{\rm
U}^2)}\,/\,{\rm I}$, expressed as a percentage, is plotted together
in Figure~\ref{fig-linpol} as a function of time, and the values are
listed in Table~\ref{tbl-linpol}.  Because Figure~\ref{fig-linpol} is
rather complex, and difficult to interpret on its own, we also include
Figure~\ref{fig-p+i} which shows separate overlaid plots of the polarized and
total intensity at each frequency.

The frequency-dependent time evolution of the P/I light curves is
reminiscent of the `expanding synchrotron bubble' model (e.g.,
\cite{h&j88}; \cite{b&v93}).  This simple model is characterized by an
optically thick rising phase in which the flux density peaks first at
the highest frequencies, followed by delayed peaks of lower intensity
at progressively lower frequencies.  The decay at all frequencies has
a power-law time dependence, with an exponent linked to the electron
energy distribution.

We have marked three apparent `bubble-type' events in the linearly
polarized ATCA data, and labelled them E1, E2 and E3 in
Figures~\ref{fig-linpol} and \ref{fig-p+i}, using the 8.6 and 9.2~GHz
curves for reference. However, it is worth noting that the beamsize of
the ATCA is larger than the angular size of \groj1655 (HR95) at all
frequencies, so we are observing effects which encompass both the
central object and the ejecta. As a result, we can only hope to
establish qualitative comparison with the sychrotron bubble model.  In
fact, contrary to the model prediction, Figure~\ref{fig-atca} shows
that the source was optically thin at TJD 9582, the time of the major
peak in flux density.

\placetable{tbl-linpol}

\placefigure{fig-linpol}

\placefigure{fig-p+i}

Figures~\ref{fig-linpol} and \ref{fig-p+i} show that the first event,
E1, is characterized by a series of declining peaks in the 8.6 and 9.2
GHz data. The peaks at 4.8 and 5.9~GHz are less well defined and their
decay is clearly delayed with respect to the higher frequencies, in
qualitative agreement with the expanding bubble model.
The delays at 1.4 and 2.3~GHz are sufficiently large that the
contributions from successive ejecta blend with one another and
produce a broad peak in both P and P/I.  It is interesting to note
that the polarized flux density (Fig.~\ref{fig-p+i}) reached
essentially the same maximum value (${\rm P}\simeq0.2$~Jy) at all
frequencies.

The early stages of event E2 also show qualitative agreement with the
expanding synchrotron bubble model, with the rise in P/I at the higher
frequencies preceding that at lower frequencies. However, the four
higher frequencies all reach a maximum together and then fall to a
minimum at the same time, behavior which is not consistent with the
model. Event E3 is also inconsistent with the model, as the fractional
polarizations at 4.8--9.2~GHz increase and decrease simultaneously.
We discuss possible interpretations of this behavior in Section
\ref{physint}.

\subsection{Position angle and rotation measure}\label{pa+rm}

Faraday rotation of the plane of polarization has been taken into
account by fitting the observed position angles (PA) to the usual
equation, ${\rm PA} = {\rm PA}_0 + ({\rm RM}) \lambda^2$, where PA$_0$
is the intrinsic position angle, RM is the rotation measure (in
rad~m$^{-2}$) and $\lambda$ is the wavelength.  The data were first
interpolated to a common epoch, namely that of the 9.2~GHz
observations.  The position angle was determined from the equation
${\rm PA}=\case{1}{2}[{\rm arctan}\,{\rm (U/Q)} + n{\pi}]$, the last
term reflecting the ambiguities of $n\pi$ that may arise at each
frequency, and then plotted against $\lambda^2$.  A weighted
least-squares linear fit yielded the intrinsic position angle
(y-intercept) and the RM (slope).  Since the position angle is an
average over the beamsize, this means that it is also an average over
the central source and the ejecta.

\placefigure{fig-pa+rm}

\placetable{tbl-parm}

Figure~\ref{fig-pa+rm} shows the temporal evolution of the intrinsic
polarization position angle and rotation measure.  The average
position angle between TJD~9579 and 9589 was $\sim 47\arcdeg$, which
is consistent with the position angle of the jets (\cite{h&r95}). This
implies that the magnetic field is perpendicular to the jets, a
phenomenon also seen in the parsec-scale relativistic jets from
powerful radio-loud AGN (e.g.  \cite{b&p84}). The overall constancy in
position angle suggests that the magnetic field is well ordered, and
maintains basically the same orientation. However, on TJD~9589.5 the
position angle drops to $\sim 26\arcdeg$, and after a brief increase
to $\sim 43\arcdeg$, remains at $\sim 24\arcdeg \pm 4\arcdeg$ during
the last three observing epochs.

The rotation measure increases rapidly from $\sim 6$ rad m$^{-2}$ on TJD~9579.5 
to $\sim 104$ rad m$^{-2}$ on TJD~$\sim 9582$. It then levels off at $\sim 60$
rad m$^{-2}$ for the rest of the observing period, the only exception
being an increase to $\sim 80$ rad m$^{-2}$ on TJD~9598.  Its relative
constancy after TJD~9582 suggests that $\sim 60$ rad m$^{-2}$ arises
from the path through the interstellar medium, while the rapid
increase prior to TJD~9582 probably reflects local effects arising at
or near the source.  The low values of rotation measure around
TJD~9579.5 imply that the magnetic field at the source is in the
opposite direction to that of the interstellar medium, and the
increase to $\sim 104$ rad m$^{-2}$ indicates that the orientation of
the magnetic field has changed direction and is now more-or-less
aligned with that in the interstellar medium.

\placefigure{fig-paplots}

While Faraday effects predict that the rotation of the plane of
polarization is proportional to $\lambda^2$, the validity of the
$\lambda^2$ dependence has been questioned by O'Dea (1989).  He
presents polarization observations of fifteen core-dominated quasars,
arguing that poor fits to the PA vs. $\lambda^2$ data for six of the
quasars cannot be due to large apparent RM values combined with
ambiguities of $n\pi$, but rather can be accounted for by
wavelength-dependent polarization structure.  A similar situation may
well apply to \groj1655.  The upper panel of Figure~\ref{fig-paplots}
shows a least-squares fit to the polarization data on days TJD~9588.60
and 9590.53, which resulted in ${\rm PA}_0=47.5^{\circ}$, ${\rm
RM}=56.4$ rad~m$^{-2}$ and ${\rm PA}_0=38.2^{\circ}$, ${\rm RM}=63.5$
rad~m$^{-2}$ respectively.  Both fits appear secure.

In contrast, the lower panel of Figure~\ref{fig-paplots} shows an
example of PA vs. $\lambda^2$ for \groj1655 on TJD~9589.53, where
$\pi$ has been added to both the 1.4 and 2.3~GHz data points (open
circles). A least-squares fit to the data (solid line) yielded ${\rm
PA}_0=26.1^{\circ}$, ${\rm RM}=65.5$~rad~m$^{-2}$, and a large reduced
$\tilde{\chi}^2$ of 12.8 for 4 d.o.f.  We tried to accommodate the
high frequency points by adding $4\pi$ and $2\pi$ to the 1.4 and
2.3~GHz data respectively (filled circles) and obtained ${\rm
PA}_0=-5.8^{\circ}$, ${\rm RM}=282.9$ rad~m$^{-2}$ (dotted line) and
an improved $\tilde{\chi}^2$ of 5.3.  However, these fitted parameters
are so inconsistent with the values obtained on TJD~9588.60 and
9590.53 (with $\tilde{\chi}^2$ values of 0.31 and 0.05 respectively)
that they seem implausible.  Hence, we conclude that the discrepancies
seen in the high frequency points on TJD~9589.53, especially those at
8.6 and 9.2~GHz, are due to wavelength dependent polarization
structure in the source.  Considering that \groj1655 is a complex
time-variable source, and that the observations are integrations over
multiple evolving ejecta, a poor fit to a simple $\lambda^2$
dependence is perhaps to be expected.

\section{Spectra}\label{spectralstuff}

\subsection{Spectral indices}\label{specind}

To fully investigate the evolution of the radio spectrum, two sets of
spectral indices were considered: the first set covers the entire
MOST/HartRAO monitoring period, from TJD~$\sim 9580$--9618, while the
second set concentrates on the ATCA data which, due to better
sampling, show more detail but for a shorter period of time.
Two-point spectral indices $\alpha$ were calculated, assuming spectra
of the form $S_{\nu}\propto\nu^{\alpha}$.  To examine the spectral
evolution more closely, low- and high-frequency spectral indices were
calculated separately.  For the MOST/HartRAO data these consisted of
the 843~MHz--5~GHz ($\alpha_{0.8-5}$) and 5--8.58~GHz
($\alpha_{5-8.5}$) bands, while for the ATCA observations the data
were divided into the 1.4--2.3~GHz ($\alpha_{1.4-2.3}$) and
4.8--8.6~GHz ($\alpha_{4.8-8.6}$) bands. In addition, a `global'
spectral index was obtained from the MOST 843~MHz and HartRAO 8.58~GHz
data ($\alpha_{0.8-8.5}$) and the MOST and ATCA 9.2~GHz data
($\alpha_{0.8-9.2}$).  

The single-dish HartRAO flux densities are less accurate than those
from the ATCA (and the VLA) and will be affected by confusing sources
in the beam; an image of the field of \groj1655 is shown in Hunstead
et al. (1997).  On the other hand, the HartRAO data span a wider time
interval than the ATCA and fill in vital gaps in the VLA light curves.
Furthermore, while confusion will affect the spectral indices
systematically, trends with time will be preserved.

\subsubsection{MOST and HartRAO}

The lower panel in Figure~\ref{fig-most+rao} shows the spectral
indices from the MOST and HartRAO data plotted as a function of time.
The MOST and HartRAO 8.58~GHz data points were interpolated to the
HartRAO 5 GHz epochs.  Between TJD~9580--9593 and TJD~9610.5--9616
only the global $\alpha_{0.8-8.5}$ spectral index is plotted, while
from TJD 9593 to 9610, $\alpha_{0.8-5}$ and $\alpha_{5-8.5}$ are
plotted separately. Prior to TJD~9593, 
$\alpha_{0.8-8.5}$ ranges between $-0.6$ and $-0.4$, whereas after
TJD~9610 the spectrum is slightly flatter overall with
$\alpha_{0.8-8.5} \approx -0.3$.  At the time of the {\sc vla3}
ejection event (TJD $\sim9596$), the 5--8.58~GHz spectrum inverts to
$\alpha_{5-8.5}\approx +0.25$ (with a large uncertainty); this
mini-outburst is also seen in the VLA data (\cite{h&r95}), most
notably at 15 and 22 GHz.  The 5--8.58~GHz spectral index then returns
to the pre-outburst value at TJD $\sim 9600$.  A second
outburst, also recorded by the VLA, occurs a few days later, with
$\alpha_{5-8.5}$ peaking at $\sim +0.4$ at TJD~$\sim 9607$ 
and steepening soon thereafter.  The inversion of the
spectrum, indicating the emergence of an optically thick component, is
consistent with the {\sc vlba2} ejection event; the precise time
correspondence is, however, uncertain because neither the HartRAO nor
VLA dataset covers the critical interval TJD 9601--9606.

\subsubsection{MOST and ATCA}\label{most_atca_specind}

\placefigure{fig-specind}

Figure~\ref{fig-specind} shows the spectral indices obtained from the
MOST/ATCA datasets plotted as a function of time.  To calculate the
843~MHz--9.2~GHz index the MOST data points were interpolated to the
9.2~GHz epochs.  The temporal evolution of the three datasets is
similar, with $\alpha_{0.8-9.2}$ and $\alpha_{1.4-2.3}$ ranging from
$\sim -0.4$ to $\sim -0.6$, while $\alpha_{4.8-8.6}$ (excluding the
last two points at TJD~9596 and 9598) varies between $-0.5$ and
$-0.8$, indicating that the high-frequency portion of the spectrum is
consistently steeper. There is an overall steepening of the spectrum
between TJD~9579 and 9582, as the source becomes optically thin at the
lower frequencies.  Immediately after this, the high-frequency
spectral index begins to flatten and is followed by the lower
frequencies on TJD~$\sim 9583$; the light curves in Figure~\ref{fig-atca} 
shows a clear dip and rise at this epoch.  This
behavior is consistent with the ejection of an optically thick
synchrotron component and points to an ejection epoch TJD~$\sim 9582$,
somewhat earlier than inferred for the {\sc vla2} event
(\cite{h&r95}).  The spectra are flattest on TJD~$\sim 9586$, after
which the spectrum steepens.  Between TJD~9596.33 and 9598.07,
$\alpha_{4.8-8.6}$ jumps from $-$0.87 to $-$0.21, coincident with the
{\sc vla3} ejection episode; the corresponding increase in the 8.6~GHz
flux density can be seen in Figure~\ref{fig-atca}.

\subsection{Linear polarization spectra}\label{linpolspectra}

\placefigure{fig-spectra}

In Figure~\ref{fig-spectra} we show a montage of overlaid plots of the
total (I) and polarized (P) intensity spectra of \groj1655.  To
construct the spectra, the flux densities at all frequencies,
including 843~MHz (but without polarization), were interpolated to the
epochs of the ATCA 9.2~GHz observations.  The P spectra, in
particular, are especially valuable as diagnostics of ejection events
and their evolution with time.  We now discuss these spectra in
conjunction with the overlaid fractional polarization plot in Figure~\ref{fig-linpol} 
and the light curves in Figure~\ref{fig-atca}.

On the first day, TJD 9579.6 (four observations spanning two hours),
the P spectrum is strongly inverted and the I spectrum shows spectral
flattening at the low frequencies.  On the following day TJD~9580.50,
the I spectrum has become essentially a power law.  The P spectrum
shows a marked increase at 2.3~GHz but is still optically thick at 1.4
GHz. By TJD~9581.64, near the peak in total intensity, the P spectrum
has flattened, with signs of an upturn at the highest frequencies
signalling another ejection (E1).  The following four P spectra,
spanning TJD~9582.17--9583.59, trace a steady increase in the
polarized flux density at the lower frequencies, a pattern which is
qualitatively consistent with the expansion of the
synchrotron-emitting region(s).

At TJD~9584.63 the P spectrum is still not completely transparent at
the lower frequencies but has steepened markedly at the higher
frequencies, with signs of a new component (E2) appearing at 8.6 and
9.2 GHz.  The development of this new component over the next three
days is again consistent with the expected trend for an expanding,
optically-thick synchrotron bubble.  Over this same time interval the
low frequency P spectrum has become optically thin.  Between TJD
9587.62 and 9588.62 we interpret the persistently high level of P for
frequencies $\geq 4.8$~GHz as indicating yet another ejection event,
even though there is no perceptible change in the total flux density.

From TJD 9588.62 to 9589.55 there is an interesting transition.  At
all frequencies the total flux density from TJD 9589.55--9591.37 lies
well above an extrapolation of the exponential decay seen over the
previous four days (Fig.\ \ref{fig-atca}), implying that a
flat-spectrum component has been introduced.  A simultaneous flux
density increase across the spectrum is {\it not\/} a characteristic
of the synchrotron bubble model, so an alternative interpretation must
be sought.

Accompanying the flux density increase at TJD 9589.55 is a sharp
decrease in the high-frequency polarized flux density and fractional
polarization which occurs simultaneously at the four frequencies $\geq
4.8$~GHz.  The high-frequency polarized flux `reappears' on TJD
9590.55 and 9591.37 but is quenched again in the same fashion two days
later on TJD 9593.36.  We put forward a speculative but
self-consistent explanation of this behavior in Section \ref{physint}.

\section{Comparison with the VLA/VLBA}\label{compare}

In Sections \ref{lightcurves}, \ref{linpol} and \ref{spectralstuff} we
have described various aspects of the radio behavior of \groj1655 as
observed with the MOST, ATCA and HartRAO. We now discuss this behavior
in the light of the images and light curves recorded with the VLA and
VLBA (\cite{h&r95}).

The light curves presented here (Figs. 1--3) are completely
consistent with those produced at the VLA (HR95), with an outburst
peaking on ${\rm TJD}\sim 9582$ followed by a more-or-less exponential
decay.  Due to sparser temporal coverage the VLA did not record the
two short duration outbursts clearly observed with MOST (S$_1$ and
S$_2$) and HartRAO (S$_2$ only).  

As discussed in Section \ref{linpolspectra}, the ATCA linear
polarization spectra (Fig.~\ref{fig-spectra}) seem to give the
clearest indication of plasmon ejection and expansion.  The E1 event
is probably composed of several short-lived contributions, beginning
at ${\rm TJD} \sim 9580$, and possibly associated with ejection event
{\sc vla1} on TJD 9578.  The {\sc vla2} event (\cite{h&r95}),
originating at TJD 9584, is not well defined because of the 6-day gap
in VLA coverage from TJD 9583 to 9589.  It was noted previously
(Section \ref{most_atca_specind}) that the overall spectral indices
pointed towards a major outburst beginning near TJD 9582.  This
interpretation is supported by our polarization spectra which suggest
that events beginning near TJD 9581.6 and 9584.6 may be confusing the
interpretation of the component positions recorded later by the VLA.

Both the {\sc vla3} and {\sc vlba2} events are well traced and
supported by our spectral index plots in Figure \ref{fig-most+rao} and
Figure \ref{fig-specind}.  It is worth noting, however, that the
steepness of the high-frequency P and I spectra in the last panel in
Figure~\ref{fig-spectra} suggests that the {\sc vla3} event began {\it
after\/} TJD 9596.32.


\section{Physical Interpretation}\label{physint} 

The radio emission from candidate black hole binaries is usually
attributed to optically thin synchrotron emission or thermal free-free
emission. The former explains the steep high frequency part of the
power-law spectrum, while the latter is invoked if the spectrum is
flat, extending to high frequency.

\subsection{Polarization and the Synchrotron Bubble Model}\label{synchbub} 

In the simple synchrotron bubble model (e.g.\ Hjellming \& Johnston
1988; Ball \& Vlassis 1993) the emission region is a homogeneous
spherical, expanding plasma cloud, composed of relativistic electrons
and embedded in a magnetic field. As the cloud (bubble) expands
adiabatically, the synchrotron emission from it becomes optically thin
at progressively lower frequencies.  A natural consequence, therefore,
is that the total intensity I and fractional polarization P/I will
peak first at high frequency and later, with smaller amplitude, at
lower frequencies.

How well does this model describe the observed properties of \groj1655?
To begin with, the radio
emission is strongly polarized, giving direct evidence for the
presence of synchrotron emission (Fig.~\ref{fig-linpol}).  During
event E1, P/I peaks at 9.2 and 8.6~GHz ahead of 5.9~GHz and 4.8~GHz.
In addition, the maximum values of P/I at the high frequencies are
larger than those at the lower frequencies, the only exception being
the 1.4~GHz P/I which is explained as a blend of contributions from
several outbursts.  Event E2 displays a similar pattern.  While these
characteristics are in {\it qualitative\/} agreement with the
synchrotron bubble model, after event E2 the P/I values at
4.8--9.2~GHz reached a minimum at almost the same time, which is not
predicted.  Thereafter, during event E3, P and P/I rise and fall
simultaneously at 4.8--9.2~GHz, again contrary to model predictions.
The polarization spectra (Section \ref{linpolspectra}, Figure~\ref{fig-spectra}) 
paint a similar picture.  Lack of compliance with
the synchrotron bubble model is perhaps not surprising considering the
complex collimated structures revealed by the VLBA observations (HR95)
and the simplicity of the model itself.

\subsection{``Core-Lobe'' Model }\label{l+c} 

As polarization is detected, we cannot dismiss the contribution from
synchrotron emission. However, we have shown that the synchrotron
bubble model cannot provide a satisfactory explanation for the time
evolution of the total intensity spectrum or the polarization in \groj1655.  
As a modification, we propose a new model which takes
into account the fact that VLBI observations of GRO J1655$-$40 have
shown multiple, time-varying emission regions (HR 95, Tingay et al.\
1995).  A hybrid model which brings in other emission mechanisms is
therefore worth exploring for this very complex system.

There are two examples in the literature where multiple components
have been invoked to explain the light curves and/or polarization:
V404 Cyg (\cite{han}) and GRS\,1915$+$105 (Fender et al. 1999).

\begin{itemize}

\item The X-ray transient, V404 Cyg (\cite{han}), shows time-dependent
polarization properties similar to those seen in \groj1655, including
little variation in polarization position angle over the 50-day
monitoring period.  However, in contrast to GRO~J1655$-$40, V404 Cyg
showed a rapid initial decay, in which the spectrum was
optically-thin, followed by a much slower decline during which the
radio spectrum remained flat or inverted.  Han \& Hjellming (1992)
interpret this behavior as requiring at least two radiating
components.

\item
During a series of major ejection events in the superluminal jet X-ray
binary, GRS\,1915$+$105, recorded at high angular resolution with
MERLIN, Fender et al.\ (1999) reported that the (stationary) core did
not show any significant polarization, even though the approaching
component was significantly polarized.  They note that the core
emission was dominated by flat spectrum oscillations and attribute the
lack of polarization to either the superposition of multiple
components with different polarization position angles, or to large
Faraday depolarization close to the binary system.

\end{itemize}

In our hybrid model for \groj1655 we assume that the emission regions
consist of a `core' (an extended plasma cloud surrounding the central
source) and `lobes' (the expanding plasmons or ejecta). The core
represents the ejecta when they are localized near the binary; it is
therefore dense and compact, and unresolved in the VLBA image.  For a
beamwidth $\sim$15 mas and an estimated distance of $\sim 3$~kpc (HR95),
 we estimate that the corresponding size of the
emitting region should be $\la 10^{14}$~cm.  Synchrotron emission
is inefficient in the core, because of strong self-absorption. During
the initial stage of ejection, the core is most compact and the
electron density is highest, making free-free emission a very
efficient process.  The electrons in the ejecta are presumably
accelerated by some non-thermal (magnetic) processes near the black
hole, so their energy distribution will have a non-Maxwellian
component giving rise to non-thermal free-free emission. If the
density in the core is sufficiently high to allow particle collisions,
a sub-population of electrons with a Maxwellian distribution will
result. These electrons will emit thermal free-free emission.

When the ejecta leave the immediate environment of the binary and
expand to form the lobes, which are then resolved in the radio images,
they eventually become transparent to synchrotron emission. Moreover,
when the density drops, free-free emission becomes less efficient,
leaving synchrotron emission, which is no longer self absorbed, as the
dominant mechanism for energy loss.

According to this model, events E1 and E2 can be understood as two
consecutive episodes of mass ejection. During the onset of the
ejection, the lobes and the core are not actually separable. As the
emission region is compact, the synchrotron emission is self-absorbed.
There is also a free-free component.  Hence the overall spectrum shows
a turn-over, with the critical frequency determined jointly by the
emissivities of the synchrotron and free-free processes.  When the
individual plasma bubbles begin to separate from the core and expand,
their free-free emission becomes unimportant. Moreover, they become
progressively transparent to synchrotron emission, first at the high
frequencies and then at the low. The emission from the compact core,
on the other hand, may still be dominated by free-free emission, which
is weak in comparison with the synchrotron emission from the expanding
lobes.

Free-free emission has a flat spectrum, and so any free-free emission
from the core does not contribute significantly to the low frequency
part of the overall spectrum. For most of the time, therefore, the
spectral properties and polarization at low frequency are
characterized by the synchrotron emission from the ejecta, especially
when that emission peaks.  However, if the synchrotron component has
declined substantially, or there is a substantial brightening in the
emission from the core, the spectral index and the fractional
polarization will both be affected, becoming observable first at the
high frequencies and then at the lower frequencies.  The simultaneous
occurrence of the minima in P/I between events E2 and E3, together
with the simultaneous rise and fall in P/I at the four higher
frequencies during event E3, can be readily explained by dilution of
the synchrotron emission by a varying free-free contribution from the
core.  The magnetic field disruption resulting from core-brightening
episodes may also explain the apparently anomalous polarization
position angles (Fig.\ \ref{fig-pa+rm}) at TJD 9589.55, 9593.36,
9596.32 and 9598.07.

A free-free core component can also explain the polarization spectra
for ${\rm TJD} > 9588.6$.  Consideration of the timescales evident in
Figure~\ref{fig-spectra}, and the VLBI structures (\cite{h&r95},
\cite{tin95}), tells us that the polarized emission at low frequency
comes from regions well away from the core, whereas the high-frequency
polarized emission is concentrated close to the core.  A sudden
increase in the size or electron density in the core will therefore
have a significant impact on P at high frequency and little or no
effect at low frequency, which seems to explain qualitatively what
occurs in the transition between TJD 9588.62 and 9589.55 (Fig.\
\ref{fig-spectra}).  We note that a small but significant upturn in P
at TJD 9589.55 between 8.6 and 9.2 GHz probably signals a new ejection
event. The following days, TJD 9590.55 and 9591.37, see an increase in
the polarized flux at high frequency, presumably because the ejecta
have moved outside the region of core absorption.  Interpretation of
the last two panels in Figure~\ref{fig-spectra} is complicated by the
poorer time sampling, but it is plausible that a similar series of
events follows the minimum in P and P/I at TJD 9593.36.

Although the hybrid `core-lobe' model has provided a more satisfactory
explanation for the radio emission from \groj1655 during the 1994
ejection episodes, we emphasize that the model is only qualitative at
this stage.  Further work is necessary to quantify the model, so that
fits to the data can be carried out.

\section{Summary and conclusions}\label{summary}

\begin{figure}
\plotone{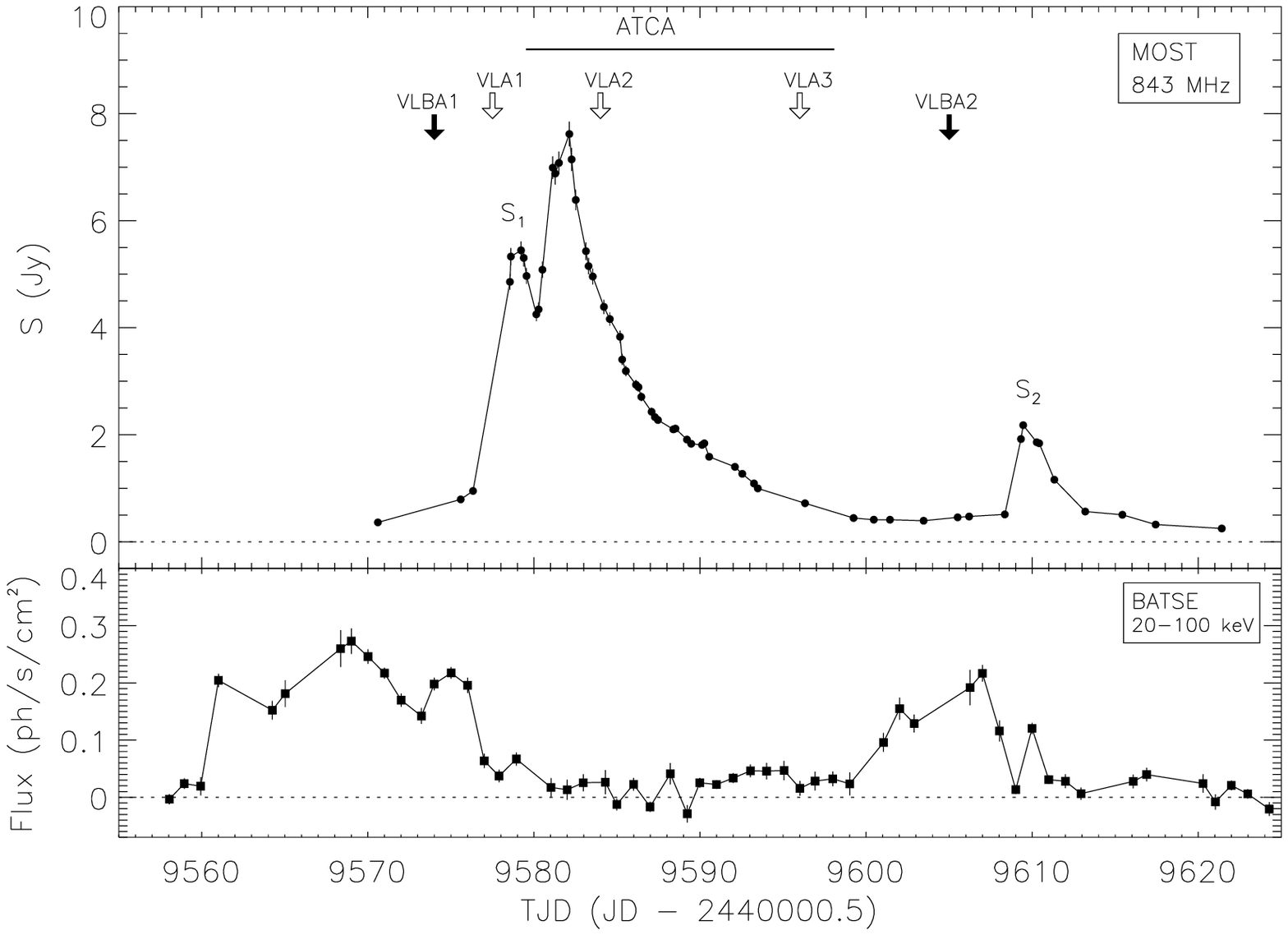}
\caption{The 1994 MOST and BATSE light curves (\cite{w&h97}).  The
open arrows mark the ejection epochs determined by \cite{h&r95} from
the VLA observations (which we label as {\sc vla1}, {\sc vla2} and
{\sc vla3}), while the filled arrows ({\sc vlba1} and {\sc vlba2})
denote their estimated VLBA ejection epochs.  S$_1$ and S$_2$ mark
short duration components discussed in the text.  The duration of the
ATCA observations is marked by a horizontal line in the upper
panel.\label{fig-most+batse}}
\end{figure}

As Figure~\ref{fig-most+batse} shows, there is a relationship between
the hard X-ray and radio emission from \groj1655. The ejection
episodes traced by the VLBA both originated during enhanced activity
in the hard X-rays, and preceded the radio outbursts recorded with the
MOST, suggesting a connection between activity near the event horizon
of the black hole and the production of relativistic electrons. One
possible explanation for the decline in intensity of the radio
outbursts with time could be that the first ejection occurred in an
environment that was relatively undisturbed, whereas the subsequent
outbursts will have taken place in an environment already disrupted by
previous activity (\cite{hjell96}).  The implications of the
X-ray/radio correlations have been discussed elsewhere (e.g.
\cite{harmon95}), with a general consensus that the hard X-rays may be
indicating enhanced accretion near the black hole which, through
processes still not well understood, triggers the formation of
relativistic radio jets.

\begin{figure}
\epsscale{1}
\plotone{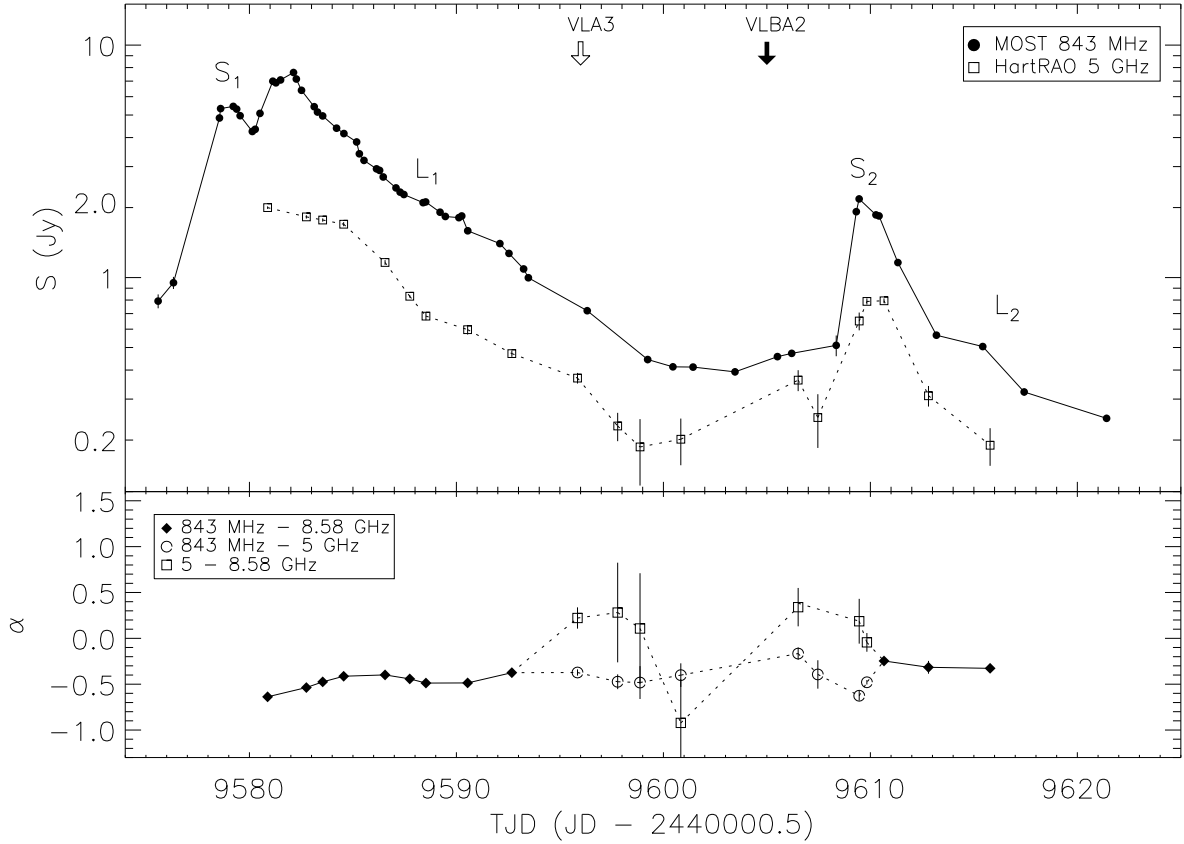}
\caption{The upper panel shows the MOST 843~MHz and HartRAO 5~GHz
light curves on a logarithmic flux density scale, with the VLA and
VLBA ejection epochs from \cite{h&r95} labelled as in Figure 1.
Labels S$_1$, S$_2$, L$_1$ and L$_2$, discussed in Section
\ref{short}, mark short and long duration components respectively.
The 843~MHz -- 8.58~GHz, 843~MHz -- 5~GHz, and 5 -- 8.58~GHz spectral
indices derived from the combined MOST and HartRAO data are plotted in
the lower panel.\label{fig-most+rao}}
\end{figure}

\begin{figure}
\epsscale{1}
\plotone{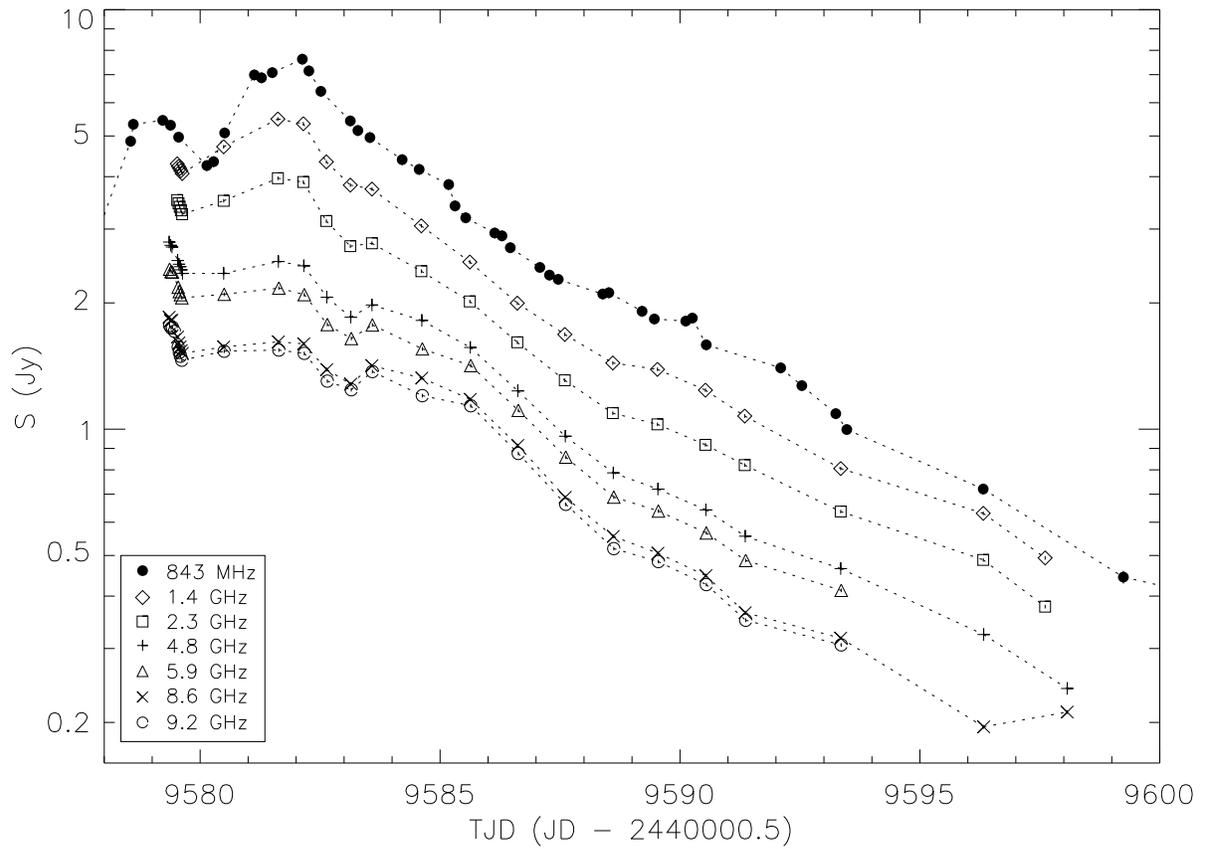}
\caption{The MOST 843~MHz and ATCA light curves shown on a logarithmic flux
density scale. The error bars lie largely within the plotting
symbols. \label{fig-atca}}
\end{figure}

\begin{figure}
\epsscale{1}
\plotone{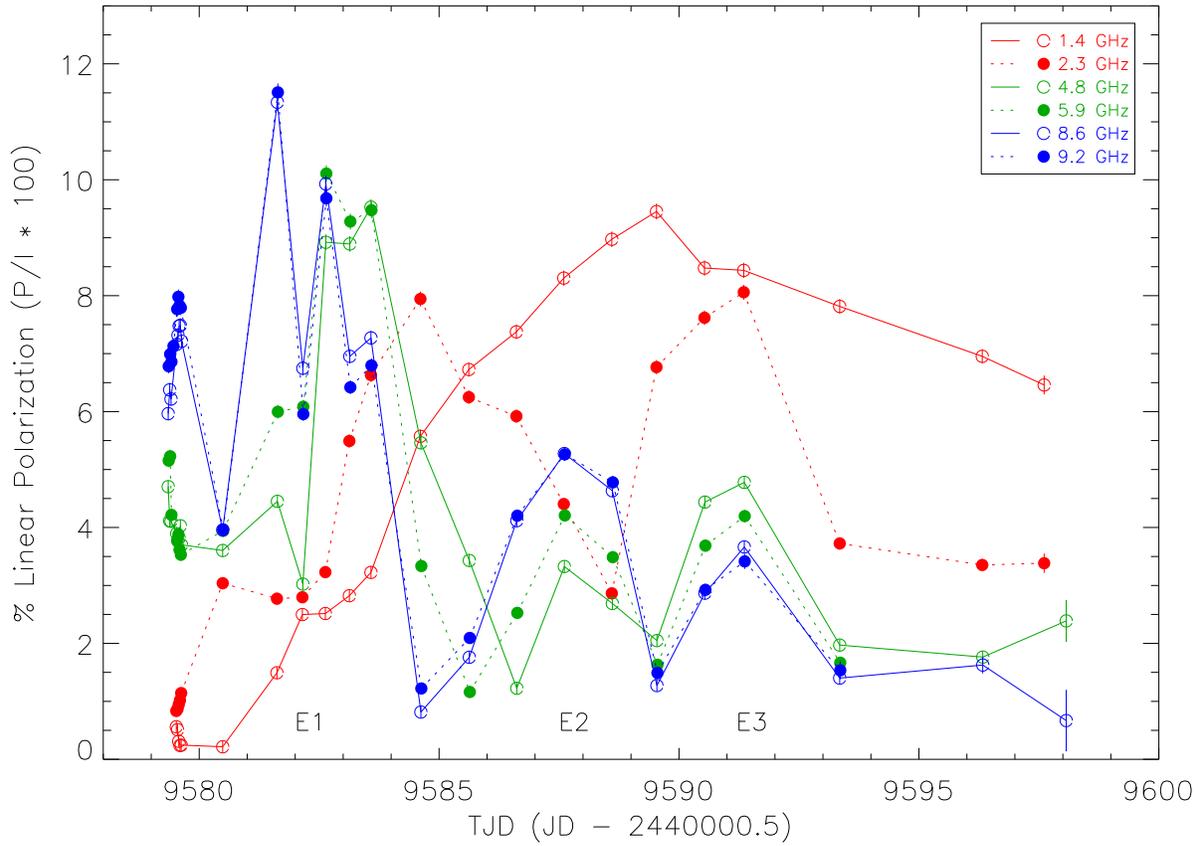}
\caption{The fractional polarization (P/I, expressed as a percentage)
at all six ATCA frequencies plotted as a function of time.  Labels E1,
E2 and E3 are used to mark specific events in polarized intensity, as
discussed in the text. \label{fig-linpol}}
\end{figure}

\begin{figure}
\epsscale{1}
\plotone{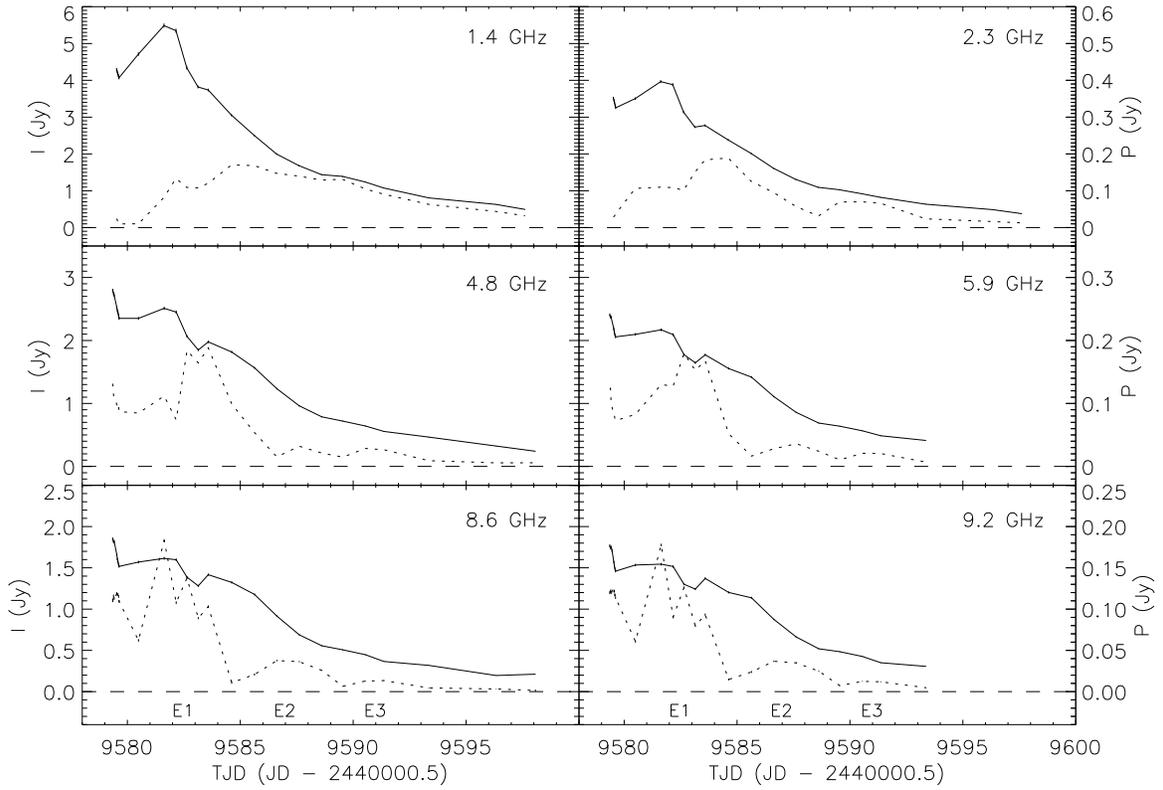}
\caption{Total (solid line) and polarized (dotted line) flux density
versus time at each of the six ATCA frequencies.  E1, E2 and E3 refer
to the same outburst events as in Figure
\ref{fig-linpol}.  Labels on the left-hand y-axes refer to the total
flux density (I), while the right-hand side labels refer to the
polarized flux density (P).
\label{fig-p+i}}
\end{figure}

\begin{figure}
\epsscale{1}
\plotone{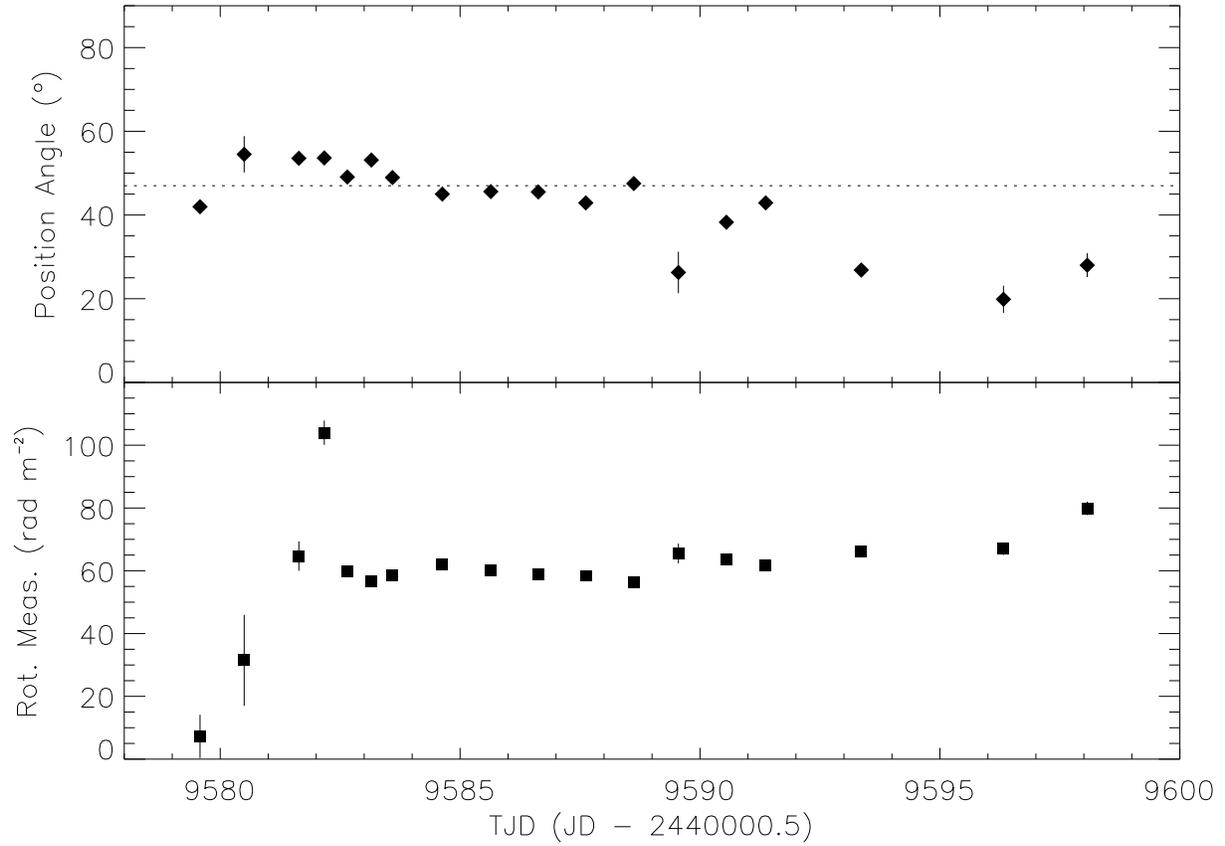}
\caption{Upper panel: the intrinsic polarization position angle
(degrees) as a function of time; the dotted line marks 47$^{\circ}$,
the position angle of the radio jets (\cite{h&r95}).  Lower panel:
rotation measure (rad m$^{-2}$) as a function of time.  The last two
points were obtained using only the 1.4, 2.3, 4.8 and 8.6~GHz
data. \label{fig-pa+rm}}
\end{figure}

\begin{figure}
\epsscale{1}
\plotone{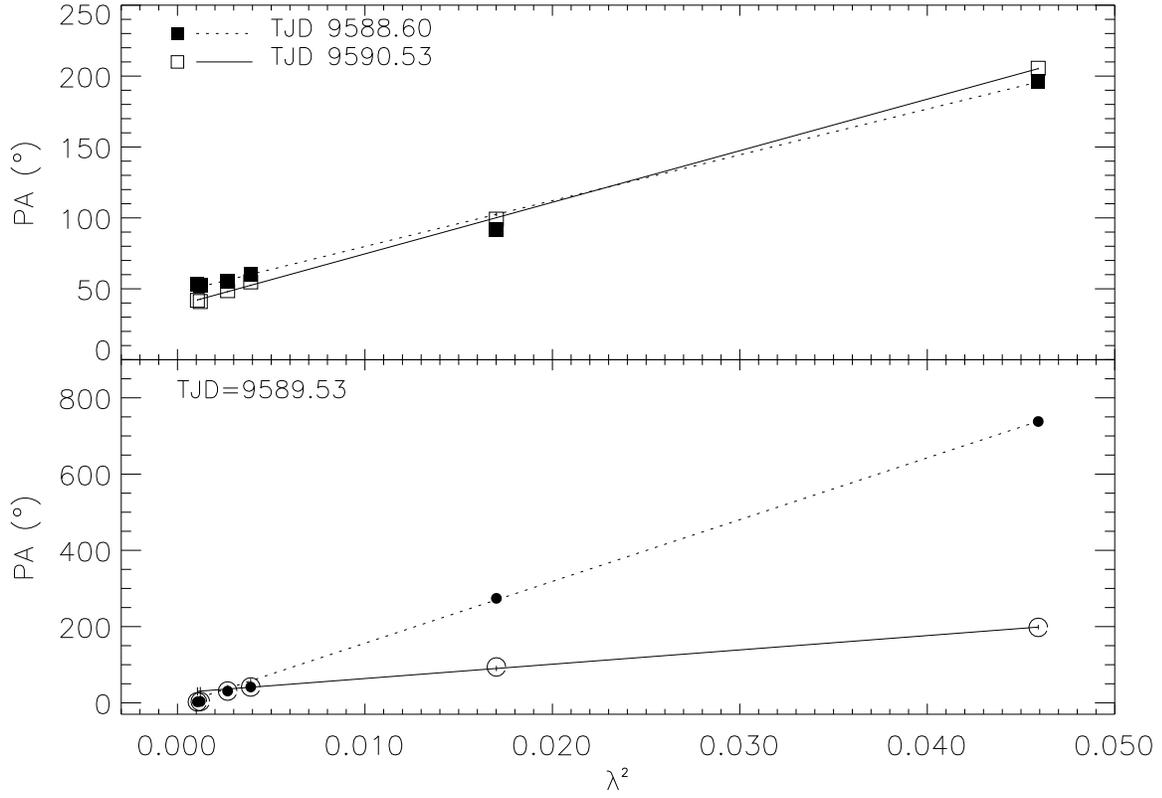}
\caption{Upper panel: PA vs. $\lambda^2$ on TJD~9588.60 and
TJD~9590.53 together with the fitted lines. Lower panel: PA
vs. $\lambda^2$ on TJD~9589.53; the open circles and solid line (the
adopted fit) show the result of adding $\pi$ to both the 1.4 and
2.3~GHz points, while the filled circles and dotted line show the
result of adding $4\pi$ to the 1.4~GHz data and $2\pi$ to the 2.3~GHz
data; see Section 4.1.  
\label{fig-paplots}}
\end{figure}

\begin{figure}
\epsscale{1}
\plotone{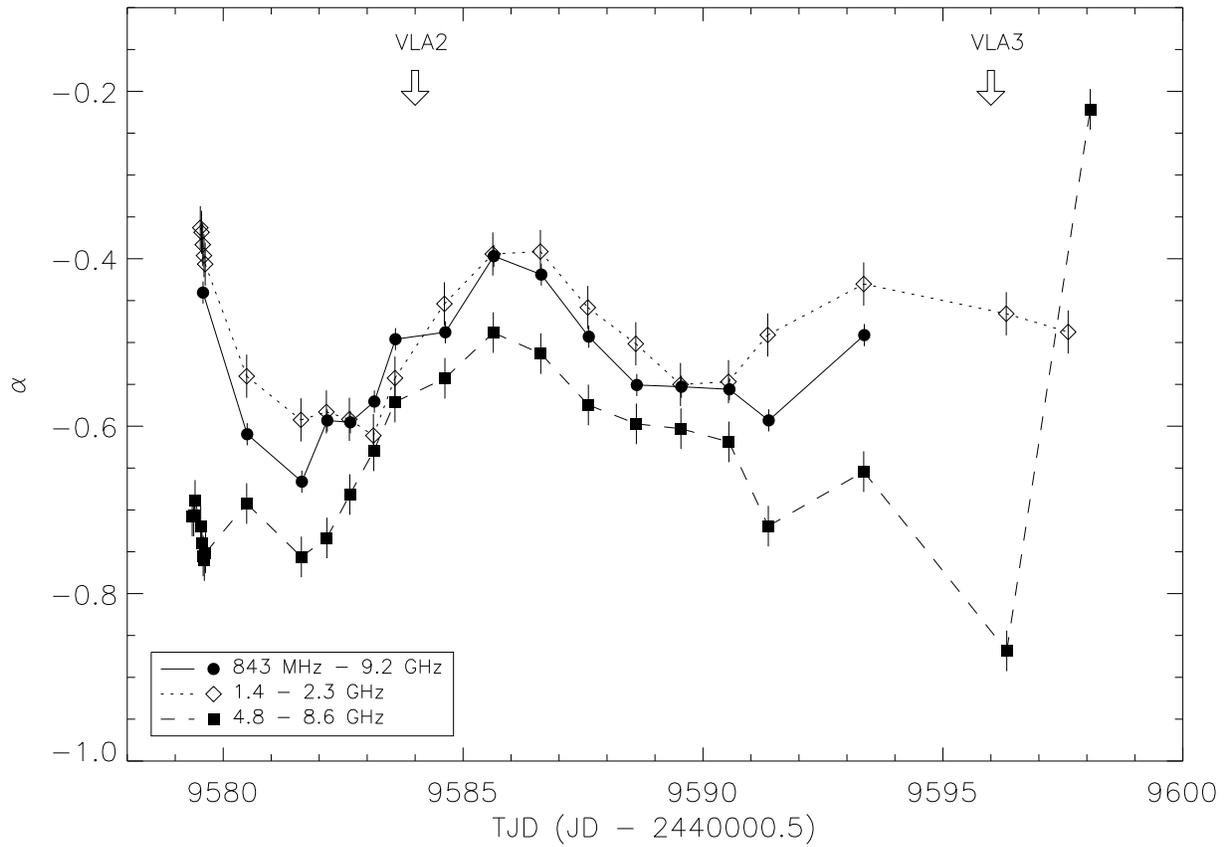}
\caption{The 843~MHz--9.2~GHz, 1.4--2.3~GHz and 4.8--8.6~GHz spectral
indices, from the combined MOST and ATCA data, plotted as a function
of time.\label{fig-specind}}
\end{figure}

\begin{figure}
\epsscale{1}
\plotone{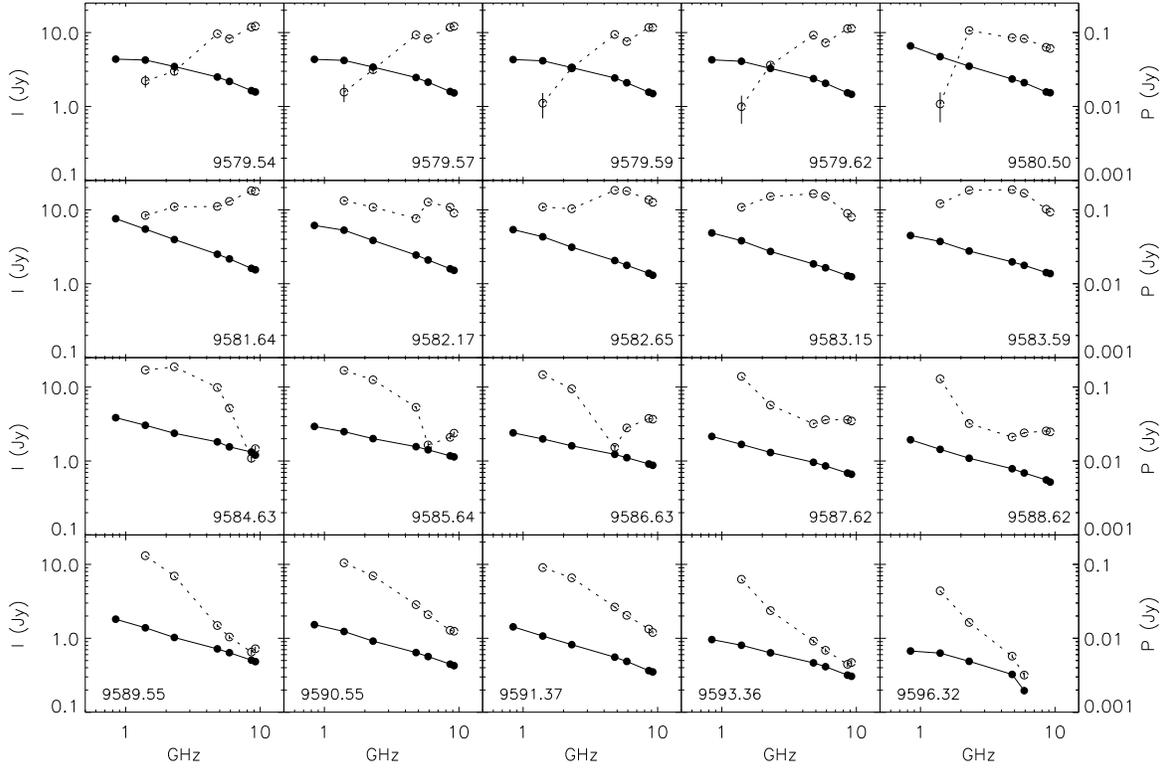}
\caption{Spectra constructed from the MOST and ATCA data.  The filled
dots and solid lines represent the combined 843~MHz--9.2~GHz MOST and
ATCA total flux density spectra (left-hand y-axis labels).  The open
circles and dashed lines represent the polarized flux from the ATCA
1.4--9.2~GHz data (right-hand y-axis labels).  The spectra on
TJD~9596.32 are constructed from the 1.4, 2.3, 4.8 and 8.6~GHz data
points alone (in addition to the MOST data for the total flux density
spectrum).  The observing epoch is marked in each
plot. \label{fig-spectra}}
\end{figure}

In general the ATCA, MOST and HartRAO flux density light curves agree
well with the VLA light curves and ejection epochs reported in
HR95. However, the better time sampling in Figures \ref{fig-most+rao}
and \ref{fig-atca} reveals the presence of short-lived events that
were not recorded by the VLA or VLBA, and shows that the light curves
do not decay as smooth exponentials.  The radio spectra, especially
the linear polarization spectra, have proved to be valuable
diagnostics of the timing of plasmon ejection events and their
subsequent evolution.

The ATCA polarization data show that the magnetic field is aligned at
right angles to the radio jets (jet ${\rm PA} = 47 \pm 1^{\circ}$), except
towards the end of the monitoring period when core contributions may
have become important.  The rotation measure initially shows a
contribution local to the \groj1655 system, but after TJD 9582.5, the
rotation measure is roughly constant at $\sim$60~rad~m$^{-2}$ which
must correspond to the interstellar value.

After examining the time evolution of the total and polarized flux
density of \groj1655 we conclude that there are specific aspects of
the behavior which cannot be explained by the simple
synchrotron bubble model.  We therefore invoke a hybrid `core-lobe'
model, with a core which emits by non-thermal (or maybe thermal)
free-free emission and lobes which are classical synchrotron
emitters.  We suggest that a similar model may apply to the other
Galactic superluminal jet X-ray binary, GRS 1915+105.

\acknowledgments

The Australia Telescope Compact Array is funded by the Commonwealth of
Australia for operation as a National Facility managed by CSIRO.  MOST
is operated by the University of Sydney and funded by grants from the
Australian Research Council.  KW acknowledges the support of the
Australian Research Council through an Australian Research Fellowship.
DM acknowledges support for his research by the European Union under
contract FMGECT950012, and thanks N.P.F.\ McKay and M.J.\ Kesteven for
assisting with the ATCA observations.  DH acknowledges financial
support from the Academy of Finland.  The director of HartRAO,
G. Nicolson, is thanked for the allocation of observing time that made
this project possible. We thank T.\ Ye for assistance with the MOST
data reduction, and J-P Macquart for his invaluable help in checking
some of the position angle fits.

\clearpage

\clearpage
 
\begin{deluxetable}{lll}
\scriptsize
\tablecaption{Journal of MOST observations. \label{tbl-most}}
\tablewidth{0pt}
\tablehead{
\colhead{UT Date} & \colhead{JD $-$ 2,440,000.5\tablenotemark{a}}   & \colhead{ S$_{0.843}$\tablenotemark{b}}  \\
\colhead{} & \colhead{} & \colhead{(Jy)}
}
\startdata 
1994 Aug 6  &      \hspace{0.4cm} 9570.60  &   0.362  $\pm$ 0.020\nl
1994 Aug 11 &      \hspace{0.4cm} 9575.59  &   0.792 $\pm$ 0.054 \nl
1994 Aug 12 &      \hspace{0.4cm} 9576.33  &   0.950 $\pm$ 0.057$^{\ast}$ \nl
1994 Aug 14 &      \hspace{0.4cm} 9578.55  &   4.86 $\pm$ 0.15 \nl
           &       \hspace{0.4cm} 9578.61  &   5.33 $\pm$ 0.16  \nl
1994 Aug 15 &      \hspace{0.4cm} 9579.22  &   5.45 $\pm$ 0.16$^{\ast}$ \nl
            &      \hspace{0.4cm} 9579.38  &   5.30 $\pm$ 0.16  \nl
            &      \hspace{0.4cm} 9579.55  &   4.97 $\pm$ 0.15  \nl
1994 Aug 16 &      \hspace{0.4cm} 9580.14  &   4.25  $\pm$ 0.13$^{\ast}$  \nl
            &      \hspace{0.4cm} 9580.28  &   4.34 $\pm$ 0.13  \nl
            &      \hspace{0.4cm} 9580.51  &   5.08 $\pm$ 0.15  \nl
1994 Aug 17 &      \hspace{0.4cm} 9581.13  &   6.99 $\pm$ 0.21$^{\ast}$ \nl
            &      \hspace{0.4cm} 9581.28  &   6.88 $\pm$ 0.21   \nl
            &      \hspace{0.4cm} 9581.50  &   7.08 $\pm$ 0.21   \nl
1994 Aug 18 &      \hspace{0.4cm} 9582.13  &   7.62 $\pm$ 0.23$^{\ast}$  \nl
            &      \hspace{0.4cm} 9582.27  &   7.14 $\pm$ 0.21   \nl
            &      \hspace{0.4cm} 9582.52  &   6.39 $\pm$ 0.19  \nl
1994 Aug 19 &      \hspace{0.4cm} 9583.13  &   5.43 $\pm$ 0.16$^{\ast}$  \nl
            &      \hspace{0.4cm} 9583.29  &   5.15 $\pm$ 0.15  \nl
            &      \hspace{0.4cm} 9583.54  &   4.96 $\pm$ 0.15 \nl
1994 Aug 20 &      \hspace{0.4cm} 9584.21  &   4.39 $\pm$ 0.13$^{\ast}$  \nl
            &      \hspace{0.4cm} 9584.56  &   4.16 $\pm$ 0.12   \nl
1994 Aug 21 &      \hspace{0.4cm} 9585.18  &   3.83 $\pm$ 0.11$^{\ast}$  \nl
            &      \hspace{0.4cm} 9585.31  &   3.41 $\pm$ 0.10   \nl
            &      \hspace{0.4cm} 9585.53  &   3.19 $\pm$ 0.10  \nl
1994 Aug 22 &      \hspace{0.4cm} 9586.14  &   2.93 $\pm$ 0.09$^{\ast}$  \nl
            &      \hspace{0.4cm} 9586.29  &   2.89 $\pm$ 0.09$^{\ast}$  \nl
            &      \hspace{0.4cm} 9586.46  &   2.71 $\pm$ 0.08   \nl
1994 Aug 23 &      \hspace{0.4cm} 9587.08  &   2.43 $\pm$ 0.07$^{\ast}$  \nl
            &      \hspace{0.4cm} 9587.28  &   2.33 $\pm$ 0.07$^{\ast}$  \nl
            &      \hspace{0.4cm} 9587.46  &   2.28 $\pm$ 0.07  \nl
1994 Aug 24 &      \hspace{0.4cm} 9588.39  &   2.10 $\pm$ 0.06$^{\ast}$ \nl
            &      \hspace{0.4cm} 9588.52  &   2.11 $\pm$ 0.06  \nl
1994 Aug 25 &      \hspace{0.4cm} 9589.21  &   1.91 $\pm$ 0.06$^{\ast}$ \nl
            &      \hspace{0.4cm} 9589.47  &   1.83 $\pm$ 0.05  \nl
1994 Aug 26 &      \hspace{0.4cm} 9590.12  &   1.81 $\pm$ 0.05$^{\ast}$ \nl
            &      \hspace{0.4cm} 9590.26  &   1.84 $\pm$ 0.06  \nl
            &      \hspace{0.4cm} 9590.55  &   1.59 $\pm$ 0.05  \nl
1994 Aug 28 &      \hspace{0.4cm} 9592.10  &   1.40 $\pm$ 0.04$^{\ast}$ \nl
            &      \hspace{0.4cm} 9592.54  &   1.27 $\pm$ 0.04$^{\ast}$ \nl
1994 Aug 29 &      \hspace{0.4cm} 9593.25  &   1.09 $\pm$ 0.03  \nl
            &      \hspace{0.4cm} 9593.48  &   0.998 $\pm$ 0.03 \nl
1994 Sept 1 &      \hspace{0.4cm} 9596.32  &   0.720 $\pm$ 0.022$^{\ast}$ \nl
1994 Sept 4 &      \hspace{0.4cm} 9599.24  &   0.444 $\pm$ 0.016  \nl
1994 Sept 5 &      \hspace{0.4cm} 9600.46  &   0.413 $\pm$ 0.013  \nl
1994 Sept 6 &      \hspace{0.4cm} 9601.43  &   0.412 $\pm$ 0.013  \nl
1994 Sept 8 &      \hspace{0.4cm} 9603.46  &   0.393 $\pm$ 0.014  \nl
1994 Sept 10 &     \hspace{0.4cm} 9605.51  &   0.457 $\pm$ 0.015 \nl
1994 Sept 11 &     \hspace{0.4cm} 9606.20  &   0.472 $\pm$ 0.015  \nl
1994 Sept 13 &     \hspace{0.4cm} 9608.35  &   0.511 $\pm$ 0.052$^{\ast}$ \nl
1994 Sept 14 &     \hspace{0.4cm} 9609.32  &   1.92 $\pm$ 0.06$^{\ast}$  \nl
             &     \hspace{0.4cm} 9609.46  &   2.18 $\pm$ 0.07  \nl
1994 Sept 15 &     \hspace{0.4cm} 9610.27  &   1.86 $\pm$ 0.06$^{\ast}$  \nl
             &     \hspace{0.4cm} 9610.42  &   1.84 $\pm$ 0.06  \nl
1994 Sept 16 &     \hspace{0.4cm} 9611.33  &   1.16 $\pm$ 0.03$^{\ast}$  \nl
1994 Sept 18 &     \hspace{0.4cm} 9613.19  &   0.565 $\pm$ 0.018 \nl
1994 Sept 20 &     \hspace{0.4cm} 9615.43  &   0.505 $\pm$ 0.016 \nl
1994 Sept 22 &     \hspace{0.4cm} 9617.43  &   0.322 $\pm$ 0.014  \nl
1994 Sept 26 &     \hspace{0.4cm} 9621.41  &   0.248 $\pm$ 0.008  \nl
1994 Sept 30 &     \hspace{0.4cm} 9625.42  &   0.151 $\pm$ 0.007  \nl
1994 Oct 3 &       \hspace{0.4cm} 9628.11  &   0.121 $\pm$ 0.006   \nl
1994 Oct 5 &       \hspace{0.4cm} 9630.38  &   0.077 $\pm$ 0.005  \nl
1994 Oct 6 &       \hspace{0.4cm} 9631.42  &   0.078 $\pm$ 0.007  \nl
1994 Oct 10&       \hspace{0.4cm} 9635.40  &   0.047 $\pm$ 0.005   \nl
1994 Oct 12 &      \hspace{0.4cm} 9637.38  &   0.040 $\pm$ 0.004   \nl
1994 Oct 18 &      \hspace{0.4cm} 9643.39  &   0.034 $\pm$ 0.006   \nl
1994 Oct 20 &      \hspace{0.4cm} 9645.21  &   0.024 $\pm$ 0.003   \nl
1994 Nov 10&       \hspace{0.4cm} 9666.37  &   0.232 $\pm$ 0.014   \nl
1994 Nov 11 &      \hspace{0.4cm} 9667.29  &   0.205 $\pm$ 0.009  \nl
1994 Nov 12 &      \hspace{0.4cm} 9668.16  &   0.116 $\pm$ 0.004   \nl
1994 Nov 13 &      \hspace{0.4cm} 9669.07  &   0.074 $\pm$ 0.002   \nl
1994 Nov 14 &      \hspace{0.4cm} 9670.32  &   0.056 $\pm$ 0.003   \nl
1994 Nov 16 &      \hspace{0.4cm} 9672.33  &   0.075 $\pm$ 0.006   \nl
\enddata
\tablenotetext{a}{At mid-observation.}
\tablenotetext{b}{Measurements made in fan-beam SCAN mode are marked with $^{\ast}$.}
\end{deluxetable}

\clearpage

\begin{deluxetable}{rcrrrrrr}
\scriptsize
\tablecaption{Journal of ATCA observations\tablenotemark{a}. \label{tbl-atca}}
\tablewidth{0pt}
\tablehead{
\colhead{UT Date} & \colhead{JD $-$ 2,440,000.5\tablenotemark{b}}   & \colhead{S$_{1.4}$}   & \colhead{S$_{2.3}$} & 
\colhead{S$_{4.8}$}  & \colhead{S$_{5.9}$} & \colhead{S$_{8.6}$} & 
\colhead{S$_{9.2}$} \\
\colhead{} & \colhead{} & \colhead{(Jy)} & \colhead{(Jy)} & \colhead{(Jy)} &
\colhead{(Jy)} & \colhead{(Jy)} & \colhead{(Jy)}
} 
\startdata 
1994 Aug 15 & 9579.35 & \nodata   & \nodata   & 2.79 & \nodata   & 1.85 & \nodata   \nl 
            & 9579.36 & \nodata   & \nodata   & \nodata   & 2.41 & \nodata   & 1.77 \nl
            & 9579.39 & \nodata   & \nodata   & 2.74 & \nodata   & 1.82 & \nodata   \nl
            & 9579.40 & \nodata   & \nodata   & \nodata   & 2.37 & \nodata   & 1.74 \nl
            & 9579.41 & \nodata   & \nodata   & 2.72 & \nodata   & 1.82 & \nodata   \nl
            & 9579.42 & \nodata   & \nodata   & \nodata   & 2.37 & \nodata   & 1.75 \nl
            & 9579.46 & \nodata   & \nodata   & \nodata   & \nodata   & \nodata   & 1.71 \nl
            & 9579.52 & 4.29 & 3.52 & \nodata   & \nodata   & \nodata   & \nodata   \nl
            & 9579.53 & \nodata   & \nodata   & 2.52 & \nodata   & 1.66 & \nodata   \nl
            & 9579.54 & \nodata   & \nodata   & \nodata   & 2.18 & \nodata   & 1.57 \nl
            & 9579.55 & 4.23 & 3.45 & \nodata   & \nodata   & \nodata   & \nodata   \nl 
            & 9579.56 & \nodata   & \nodata   & 2.47 & \nodata   & 1.61 & \nodata   \nl
            & 9579.57 & 4.18 & 3.39 & \nodata   & 2.13 & \nodata   & 1.52 \nl
            & 9579.58 & \nodata   & \nodata   & 2.44 & \nodata   & 1.57 & \nodata \nl
            & 9579.59 & \nodata   & \nodata   & \nodata   & 2.09 & \nodata   & 1.49 \nl
            & 9579.60 & 4.13 & 3.32 & \nodata   & \nodata   & \nodata   & \nodata   \nl
            & 9579.61 & \nodata   & \nodata   & 2.40 & \nodata   & 1.54 & \nodata   \nl
            & 9579.62 & 4.07 & 3.25 & \nodata   & 2.06 & \nodata   & 1.46  \nl
            & 9579.63 & \nodata   & \nodata   & 2.35 & \nodata   & 1.52 & \nodata   \nl
1994 Aug 16 & 9580.49 & 4.71 & 3.50 & 2.35 & \nodata   & 1.57 & \nodata   \nl
            & 9580.50 & \nodata   & \nodata   & \nodata   & 2.10 & \nodata   & 1.53 \nl
1994 Aug 17 & 9581.62 & 5.49 & 3.96 & \nodata   & \nodata   & \nodata   & \nodata   \nl
            & 9581.63 & \nodata   & \nodata   & 2.51 & \nodata   & 1.61 & \nodata   \nl
            & 9581.64 & \nodata   & \nodata   & \nodata   & 2.17 & \nodata   & 1.54 \nl
1994 Aug 18 & 9582.15 & 5.35 & 3.88 & \nodata      &  \nodata     & \nodata      & \nodata      \nl
            & 9582.16 & \nodata   & \nodata   & 2.45 & \nodata   & 1.60 & \nodata   \nl
            & 9582.17 & \nodata   & \nodata   & \nodata   & 2.09 & \nodata   & 1.52 \nl
            & 9582.63 & 4.34 & 3.13 & \nodata   & \nodata   & \nodata   & \nodata   \nl
            & 9582.64 & \nodata   & \nodata   & 2.06 & \nodata   & 1.39 & \nodata   \nl
            & 9582.65 & \nodata   & \nodata   & \nodata   & 1.78 & \nodata   & 1.30 \nl
1994 Aug 19 & 9583.13 & 3.82 & 2.73 & \nodata   & \nodata   & \nodata   & \nodata   \nl
            & 9583.14 & \nodata   & \nodata   & 1.85 & \nodata   & 1.28 & \nodata   \nl
            & 9583.15 & \nodata   & \nodata   & \nodata   & 1.64 & \nodata   & 1.24 \nl
            & 9583.58 & 3.74 & 2.77 & 1.98 & \nodata   & 1.42 & \nodata   \nl
            & 9583.59 & \nodata   & \nodata   & \nodata   & 1.77 & \nodata   & 1.37 \nl
1994 Aug 20 & 9584.61 & 3.05 & 2.38 & \nodata   & \nodata   & \nodata   & \nodata   \nl
            & 9584.62 & \nodata   & \nodata   & 1.82 & \nodata   & 1.32 & \nodata   \nl
            & 9584.63 & \nodata   & \nodata   & \nodata   & 1.55 & \nodata   & 1.20 \nl
1994 Aug 21 & 9585.62 & 2.50 & 2.02 & \nodata   & \nodata   & \nodata   & \nodata   \nl
            & 9585.63 & \nodata   & \nodata   & 1.57 & \nodata   & 1.18 & \nodata   \nl
            & 9585.64 & \nodata   & \nodata   & \nodata   & 1.42 & \nodata   & 1.14 \nl
1994 Aug 22 & 9586.61 & 2.00 & 1.61 & \nodata   & \nodata   & \nodata   & \nodata   \nl
            & 9586.62 & \nodata   & \nodata   & 1.23 & \nodata   & 0.915 & \nodata  \nl
            & 9586.63 & \nodata   & \nodata   & \nodata   & 1.11 & \nodata    & 0.875 \nl
1994 Aug 23 & 9587.60 & 1.68 & 1.31 & \nodata   & \nodata   & \nodata    & \nodata   \nl
            & 9587.61 & \nodata   & \nodata   & 0.962 & \nodata    & 0.683 & \nodata  \nl
            & 9587.62 & \nodata   & \nodata   & \nodata   & 0.858  & \nodata   & 0.662 \nl
1994 Aug 24 & 9588.60 & 1.44 & 1.09 & \nodata   &  \nodata    & \nodata    & \nodata  \nl
            & 9588.61 & \nodata   & \nodata   & 0.787 & \nodata    & 0.555 & \nodata  \nl
            & 9588.62 & \nodata   & \nodata   & \nodata   & 0.689  & \nodata    & 0.519 \nl
1994 Aug 25 & 9589.53 & 1.39 & 1.03 & \nodata   &  \nodata    & \nodata    & \nodata   \nl
            & 9589.54 & \nodata   &  \nodata  & 0.720 & \nodata    & 0.506 & \nodata   \nl
            & 9589.55 & \nodata   &  \nodata  & \nodata    & 0.638 & \nodata    & 0.483 \nl
1994 Aug 26 & 9590.53 & 1.24 & 0.918 & \nodata   & \nodata    & \nodata    & \nodata    \nl
            & 9590.54 & \nodata   &  \nodata  & 0.642 & \nodata    & 0.448 & \nodata  \nl
            & 9590.55 & \nodata   & \nodata   &  \nodata   & 0.566 & \nodata    & 0.426 \nl
1994 Aug 27 & 9591.35 & 1.07 & 0.821 & \nodata   & \nodata    & \nodata    & \nodata   \nl
            & 9591.36 & \nodata   & \nodata   & 0.556 & \nodata    & 0.365 & \nodata   \nl
            & 9591.37 & \nodata   &  \nodata  &  \nodata   & 0.487 & \nodata    & 0.350 \nl
1994 Aug 29 & 9593.35 & 0.806 & 0.636 & 0.465  & \nodata & 0.318   & \nodata    \nl
            & 9593.36 & \nodata    & \nodata  &  \nodata   & 0.413 & \nodata    & 0.306 \nl
1994 Sept 1 & 9596.32 & 0.630 & 0.488 & \nodata  &  \nodata   & \nodata    & \nodata    \nl
            & 9596.33 & \nodata    &  \nodata   & 0.324 & \nodata  & 0.195 & \nodata    \nl
1994 Sept 2 & 9597.61 & 0.493 & 0.378 & \nodata    & \nodata  & \nodata    & \nodata    \nl
1994 Sept 3 & 9598.07 & \nodata    & \nodata  & 0.241 & \nodata    & 0.212 & \nodata    \nl
\enddata   
\tablenotetext{a}{Flux densities are Stokes I values, with errors of 1\%.}
\tablenotetext{b}{At mid-observation.}
\end{deluxetable}

\clearpage

\begin{deluxetable}{lclcl}
\scriptsize
\tablecaption{Journal of HartRAO observations. \label{tbl-rao}}
\tablewidth{0pt}
\tablehead{
\colhead{UT Date} & \colhead{JD $-$ 2,440,000.5}   & \colhead{S$_{5}$}   & 
\colhead{JD $-$ 2,440,000.5} & \colhead{S$_{8.58}$} \\
\colhead{} & \colhead{} & \colhead{(Jy)} & \colhead{} & \colhead{(Jy)}
} 
\startdata 
1994 Aug 16 & 9580.86 & 1.88 $\pm$   0.06  & 9580.82 & 1.77 $\pm$  0.02  \nl
            & 9580.88 & 2.05 $\pm$   0.04  & 9580.86 & 1.73 $\pm$  0.12  \nl
            & 9580.90 & 2.00 $\pm$   0.04  & 9580.91 & 1.44 $\pm$  0.02 \nl
1994 Aug 18 & 9582.74 & 1.88 $\pm$   0.04  & 9582.75 & 1.57 $\pm$  0.03 \nl
            & 9582.77 & 1.77 $\pm$   0.04  & 9582.77 & 1.46 $\pm$  0.02 \nl
1994 Aug 19 & 9583.52 & 1.59 $\pm$   0.06 & 9583.52 & 1.46 $\pm$  0.02 \nl
            & 9583.53 & 1.76 $\pm$   0.05 & 9583.54 & 1.49 $\pm$  0.06  \nl
            & 9583.55 & 1.85 $\pm$   0.06 & 9583.55 & 1.68 $\pm$  0.06 \nl
            & 9583.57 & 1.82 $\pm$   0.05 & 9583.57 & 1.68 $\pm$  0.07 \nl
1994 Aug 20 & 9584.53 & 1.72 $\pm$   0.06 & 9584.54 & 1.44 $\pm$  0.05 \nl
            & 9584.55 & 1.68 $\pm$   0.05 & 9584.56 & 1.54 $\pm$  0.06 \nl
            & 9584.57 & 1.69 $\pm$   0.07 & 9584.58 & 1.52 $\pm$  0.07 \nl
1994 Aug 22 & 9586.52 & 1.14 $\pm$   0.05 & 9586.53 & 1.01 $\pm$  0.05 \nl
            & 9586.54 & 1.12 $\pm$   0.04 & 9586.55 & 0.90 $\pm$  0.03  \nl
            & 9586.56 & 1.23 $\pm$   0.06 & 9586.57 & 1.16 $\pm$  0.09 \nl
            & 9586.59 & 1.20 $\pm$   0.06 & 9586.59 & 1.15 $\pm$  0.08  \nl
1994 Aug 23 & 9587.54 & 0.91 $\pm$   0.05 & 9587.54 & 0.76 $\pm$  0.02 \nl
            & 9587.56 & 0.94 $\pm$   0.07 & 9587.64 & 0.67 $\pm$  0.02 \nl
            & 9587.63 & 0.83 $\pm$   0.04 & 9587.66 & 0.78 $\pm$  0.01 \nl
            & 9587.65 & 0.77 $\pm$   0.03 & 9587.68 & 0.84 $\pm$  0.04 \nl
            & 9587.68 & 0.91 $\pm$   0.05 & 9587.71 & 0.69 $\pm$  0.02 \nl
            & 9587.70 & 0.77 $\pm$   0.03 & 9587.73 & 0.77 $\pm$  0.07  \nl
            & 9587.72 & 0.88 $\pm$   0.03 & 9587.75 & 0.76 $\pm$  0.04 \nl
            & 9587.74 & 0.82 $\pm$   0.03 & 9587.79 & 0.76 $\pm$  0.02  \nl
            & 9587.77 & 0.88 $\pm$   0.05 & 9587.81 & 0.77 $\pm$  0.02 \nl
            & 9587.79 & 0.82 $\pm$   0.04 & 9587.84 & 0.88 $\pm$  0.10 \nl
            & 9587.80 & 0.88 $\pm$   0.06 & 9587.86 & 0.89 $\pm$  0.05 \nl
            & 9587.83 & 0.88 $\pm$   0.07 & 9587.88 & 0.93 $\pm$  0.02  \nl
            & 9587.85 & 0.85 $\pm$   0.18 & \nodata     & \nodata      \nl
            & 9587.87 & 0.77 $\pm$   0.03 & \nodata     & \nodata      \nl
            & 9587.89 & 0.93 $\pm$   0.07 & \nodata     & \nodata      \nl
            & 9587.92 & 0.91 $\pm$   0.07 &  \nodata    & \nodata      \nl
1994 Aug 24 & 9588.50 & 0.63 $\pm$   0.06 & 9588.50 & 0.68 $\pm$   0.07 \nl
            & 9588.52 & 0.67 $\pm$   0.05 & 9588.52 & 0.65 $\pm$   0.03 \nl
            & 9588.54 & 0.64 $\pm$   0.05 & 9588.55 & 0.59 $\pm$   0.06 \nl
            & 9588.56 & 0.78 $\pm$   0.05 & 9588.57 & 0.60 $\pm$   0.04 \nl
1994 Aug 26 & 9590.51 & 0.61 $\pm$   0.04 & 9590.52 & 0.52 $\pm$   0.04 \nl
            & 9590.54 & 0.56 $\pm$   0.04 & 9590.54 & 0.48 $\pm$   0.04  \nl 
            & 9590.58 & 0.65 $\pm$   0.07 & 9590.56 & 0.48 $\pm$   0.03 \nl
            &  \nodata    & \nodata                & 9590.59 & 0.52 $\pm$   0.07 \nl
1994 Aug 28 & 9592.59 & 0.32 $\pm$   0.13 & 9592.64 & 0.39 $\pm$   0.05 \nl
            & 9592.62 & 0.39 $\pm$   0.03 & 9592.68 & 0.51 $\pm$   0.07 \nl
            & 9592.64 & 0.39 $\pm$   0.04 & 9592.70 & 0.39 $\pm$   0.03 \nl
            & 9592.66 & 0.51 $\pm$   0.10 & 9592.86 & 0.38 $\pm$   0.03 \nl
            & 9592.67 & 0.53 $\pm$   0.03 & 9592.88 & 0.49 $\pm$   0.09 \nl
            & 9592.69 & 0.52 $\pm$   0.03 & \nodata     & \nodata            \nl
            & 9592.87 & 0.54 $\pm$   0.05 & \nodata     & \nodata            \nl
1994 Aug 31 & 9595.83 & 0.36 $\pm$   0.03 & 9595.83 & 0.35 $\pm$   0.03 \nl
            & 9595.85 & 0.37 $\pm$   0.03 & 9595.86 & 0.43 $\pm$   0.02 \nl
            & 9595.87 & 0.39 $\pm$   0.05 & 9595.88 & 0.49 $\pm$   0.13 \nl
1994 Sept 2 & 9597.75 & 0.23 $\pm$   0.08 & 9597.75 & 0.27 $\pm$   0.07 \nl
            & 9597.77 & 0.27 $\pm$   0.07 & \nodata     & \nodata           \nl
            & 9597.79 & 0.23 $\pm$   0.25 & \nodata     & \nodata           \nl
            & 9597.81 & 0.24 $\pm$   0.05 & \nodata     & \nodata            \nl
            & 9597.83 & 0.19 $\pm$   0.06 & \nodata     & \nodata             \nl
1994 Sept 3 & 9598.85 & 0.19 $\pm$   0.16 & 9598.83 & 0.21 $\pm$   0.01 \nl
            & 9598.87 & 0.11 $\pm$   0.10 & 9598.86 & 0.10 $\pm$   0.05 \nl
            & 9598.89 & 0.24 $\pm$   0.10 & 9598.90 & 0.12 $\pm$   0.09 \nl
1994 Sept 5 & 9600.82 & 0.16 $\pm$   0.06 & 9600.81 & 0.12 $\pm$   0.01 \nl
            & 9600.86 & 0.26 $\pm$   0.07 & \nodata     & \nodata                \nl
1994 Sept 6 &  \nodata    & \nodata                &  9601.49 & 0.19 $\pm$  0.03 \nl
            &   \nodata   & \nodata                &  9601.53 & 0.18 $\pm$  0.03 \nl
            &   \nodata   & \nodata                &  9601.56 & 0.18 $\pm$  0.03 \nl
1994 Sept 8 &   \nodata   & \nodata                &  9603.62 & 0.33 $\pm$  0.08 \nl
1994 Sept 11 & 9606.50 & 0.42 $\pm$   0.05 & 9606.47 & 0.38 $\pm$  0.06 \nl
             & 9606.52 & 0.25 $\pm$   0.06 & 9606.51 & 0.44 $\pm$  0.02 \nl
1994 Sept 12 & 9607.46 & 0.25 $\pm$   0.06 & \nodata     & \nodata           \nl
1994 Sept 14 & 9609.46 & 0.65 $\pm$   0.06 & 9609.47 & 0.72 $\pm$  0.07 \nl
             & 9609.80 & 0.69 $\pm$   0.05 & 9609.81 & 0.77 $\pm$  0.06 \nl
             & 9609.82 & 0.89 $\pm$   0.05 & 9609.85 & 0.78 $\pm$  0.04  \nl
             & 9609.84 & 0.79 $\pm$   0.04 & 9609.87 & 0.74 $\pm$  0.10 \nl
             & 9609.87 & 0.81 $\pm$   0.05 &  \nodata    & \nodata              \nl
1994 Sept 15 & 9610.61 & 0.85 $\pm$   0.03 & 9610.62 & 0.67 $\pm$  0.03 \nl
             & 9610.66 & 0.79 $\pm$   0.03 & 9610.67 & 0.74 $\pm$  0.02 \nl
             & 9610.70 & 0.74 $\pm$   0.03 & 9610.71 & 0.77 $\pm$  0.04 \nl
1994 Sept 17 & 9612.80 & 0.32 $\pm$   0.03 & 9612.80 & 0.27 $\pm$  0.04 \nl
             & 9612.82 & 0.25 $\pm$   0.08 & \nodata     & \nodata              \nl
1994 Sept 20 & 9615.79 & 0.19 $\pm$   0.03 & 9615.79 & 0.18 $\pm$  0.04 \nl
             &   \nodata   & \nodata                & 9615.81 & 0.22 $\pm$  0.01 \nl
             &   \nodata   & \nodata                & 9615.85 & 0.26 $\pm$  0.02 \nl
1994 Sept 21 &   \nodata   & \nodata                & 9616.76 & 0.11 $\pm$  0.02 \nl
             &   \nodata   & \nodata                & 9616.78 & 0.18 $\pm$  0.03 \nl
             &   \nodata   & \nodata                & 9616.80 & 0.26 $\pm$  0.02 \nl
1994 Sept 24 &   \nodata   & \nodata                & 9619.71 & 0.13 $\pm$  0.05 \nl
             &   \nodata   & \nodata                & 9619.76 & 0.15 $\pm$  0.03 \nl
1994 Oct 1   &   \nodata   & \nodata                & 9626.75 & 0.08 $\pm$ 0.04 \nl
\enddata
\end{deluxetable}

\clearpage

\begin{deluxetable}{rcrrrrrr}
\scriptsize
\tablecaption{Fractional linear polarization\tablenotemark{a}. \label{tbl-linpol}}
\tablewidth{0pt}
\tablehead{
\colhead{UT Date} & \colhead{JD $-$ 2,440,000.5\tablenotemark{b}}   & 
        \multicolumn{6}{c}{P/I $\times$ 100} \\  \cline{3-8}\\[-3mm]
 & & \colhead{1.4 GHz} & \colhead{2.3 GHz} & \colhead{4.8 GHz} & \colhead{5.9 GHz}
   & \colhead{8.6 GHz} & \colhead{9.2 GHz} 
}
\startdata 
1994 Aug 15 & 9579.35 &   \nodata       & \nodata         & 4.70$\pm$0.11 &                 & 5.96$\pm$0.12 & \nodata  \nl 
            & 9579.36 &   \nodata       & \nodata         &  \nodata       & 5.15$\pm$0.11 & \nodata         & 6.78$\pm$0.12 \nl
            & 9579.39 &   \nodata       & \nodata         & 4.11$\pm$0.11 & \nodata         & 6.37$\pm$0.12 & \nodata \nl
            & 9579.40 &   \nodata       & \nodata         & \nodata         & 5.23$\pm$0.11 &  \nodata        & 6.99$\pm$0.12 \nl
            & 9579.41 &   \nodata       & \nodata         & 4.11$\pm$0.11 &   \nodata       & 6.22$\pm$0.12 & \nodata \nl
            & 9579.42 &   \nodata       & \nodata         & \nodata         & 4.22$\pm$0.11 &  \nodata        & 6.86$\pm$0.12 \nl
            & 9579.46 &   \nodata       & \nodata         &    \nodata      &  \nodata        &   \nodata       & 7.13$\pm$0.12 \nl
            & 9579.52 & 0.561$\pm$0.100 & 0.834$\pm$0.100 & \nodata         & \nodata         & \nodata         & \nodata \nl
            & 9579.53 &   \nodata       & \nodata         & 3.89$\pm$0.11 & \nodata         & 7.16$\pm$0.12 &\nodata \nl
            & 9579.54 &   \nodata       & \nodata         &   \nodata       & 3.78$\pm$0.11 &   \nodata       & 7.77$\pm$0.13 \nl
            & 9579.55 & 0.510$\pm$0.100 & 0.870$\pm$0.100 & \nodata         & \nodata         &  \nodata        &  \nodata \nl
            & 9579.56 &   \nodata       & \nodata         & 3.76$\pm$0.11 & \nodata         & 7.32$\pm$0.12 & \nodata \nl
            & 9579.57 & 0.311$\pm$0.100 & 0.945$\pm$0.100 &   \nodata       & 3.88$\pm$0.11 &   \nodata       & 7.98$\pm$0.13  \nl
            & 9579.58 &   \nodata       & \nodata         & 3.82$\pm$0.11 & \nodata         & 7.48$\pm$0.12 &  \nodata\nl
            & 9579.59 &   \nodata       & \nodata         &  \nodata        & 3.62$\pm$0.11 &    \nodata      & 7.83$\pm$0.13 \nl
            & 9579.60 & 0.239$\pm$0.100 & 1.02$\pm$0.10 & \nodata         & \nodata         & \nodata         &  \nodata \nl
            & 9579.61 &   \nodata       & \nodata         & 4.03$\pm$0.11 &  \nodata        & 7.49$\pm$0.12 &\nodata  \nl
            & 9579.62 & 0.244$\pm$0.100 & 1.14$\pm$0.10 &   \nodata       & 3.53$\pm$0.11 &   \nodata       & 7.79$\pm$0.13  \nl
            & 9579.63 &   \nodata       & \nodata         & 3.70$\pm$0.11 &  \nodata        & 7.21$\pm$0.12 & \nodata \nl
1994 Aug 16 & 9580.49 & 0.216$\pm$0.100 & 3.04$\pm$0.10   & 3.60$\pm$0.01 & \nodata         & 3.96$\pm$0.11 & \nodata \nl
            & 9580.50 &   \nodata       & \nodata         & \nodata         & 3.94$\pm$0.11 & \nodata         & 3.96$\pm$0.11 \nl
1994 Aug 17 & 9581.62 & 1.49$\pm$0.10 & 2.77$\pm$0.10 & \nodata         & \nodata         & \nodata         & \nodata  \nl
            & 9581.63 &   \nodata       & \nodata         & 4.45$\pm$0.11 & \nodata         & 11.34$\pm$0.15 & \nodata  \nl
            & 9581.64 &   \nodata       & \nodata         &  \nodata        & 6.00$\pm$0.12 & \nodata         & 11.51$\pm$0.15 \nl
1994 Aug 18 & 9582.15 & 2.50$\pm$0.10 & 2.80$\pm$0.10 & \nodata         & \nodata         & \nodata         & \nodata  \nl
            & 9582.16 &   \nodata       & \nodata         & 3.02$\pm$0.10 & \nodata         & 6.75$\pm$0.12 & \nodata \nl
            & 9582.17 &   \nodata       & \nodata         & \nodata         & 6.08$\pm$0.12 &   \nodata       & 5.96$\pm$0.12 \nl
            & 9582.63 & 2.51$\pm$0.10 & 3.23$\pm$0.10 &\nodata          & \nodata         & \nodata         & \nodata   \nl
            & 9582.64 &    \nodata      & \nodata         & 8.92$\pm$0.13 &  \nodata        & 9.93$\pm$0.14 & \nodata \nl
            & 9582.65 &    \nodata      & \nodata         &   \nodata       & 10.11$\pm$0.14 &    \nodata     & 9.68$\pm$0.14 \nl
1994 Aug 19 & 9583.13 & 2.82$\pm$0.10 & 5.49$\pm$0.11 & \nodata         & \nodata         & \nodata         & \nodata \nl
            & 9583.14 &    \nodata      & \nodata         & 8.89$\pm$0.13 &    \nodata      & 6.95$\pm$0.12 & \nodata  \nl
            & 9583.15 &    \nodata      & \nodata         &   \nodata       & 9.28$\pm$0.14 &     \nodata     & 6.42$\pm$0.12  \nl
            & 9583.58 & 3.22$\pm$0.10 & 6.63$\pm$0.12     & 9.53$\pm$0.14 &  \nodata        & 7.27$\pm$0.12 & \nodata \nl
            & 9583.59 &      \nodata    & \nodata         &     \nodata     & 9.48$\pm$0.14 &   \nodata       & 6.80$\pm$0.12 \nl
1994 Aug 20 & 9584.61 & 5.57$\pm$0.11 & 7.94$\pm$0.13 & \nodata         &\nodata          & \nodata         &  \nodata \nl
            & 9584.62 &      \nodata    & \nodata         & 5.46$\pm$0.11 & \nodata         & 0.815$\pm$0.100 & \nodata \nl
            & 9584.63 &      \nodata    & \nodata         &  \nodata        & 3.33$\pm$0.10 & \nodata         & 1.22$\pm$0.10 \nl
1994 Aug 21 & 9585.62 & 6.73$\pm$0.12 & 6.25$\pm$0.12 & \nodata         &  \nodata        & \nodata         & \nodata \nl
            & 9585.63 &       \nodata   & \nodata         & 3.43$\pm$0.11 & \nodata         & 1.76$\pm$0.10 & \nodata   \nl
            & 9585.64 &       \nodata   & \nodata         &  \nodata        & 1.16$\pm$0.10 &  \nodata        & 2.09$\pm$0.10 \nl
1994 Aug 22 & 9586.61 & 7.38$\pm$0.12 & 5.92$\pm$0.12 &\nodata          &\nodata          & \nodata          & \nodata \nl
            & 9586.62 &        \nodata  & \nodata         & 1.22$\pm$0.10 &   \nodata       & 4.11$\pm$0.11 & \nodata \nl
            & 9586.63 &        \nodata  & \nodata         &   \nodata       & 2.53$\pm$0.10 &       \nodata   & 4.20$\pm$0.11 \nl
1994 Aug 23 & 9587.60 & 8.30$\pm$0.13 & 4.40$\pm$0.11 & \nodata         & \nodata         & \nodata          & \nodata \nl
            & 9587.61 &        \nodata  & \nodata         & 3.33$\pm$0.10 &   \nodata       & 5.28$\pm$0.11 & \nodata \nl
            & 9587.62 &         \nodata & \nodata         &   \nodata       & 4.21$\pm$0.11 &  \nodata        & 5.26$\pm$0.11 \nl
1994 Aug 24 & 9588.60 & 8.97$\pm$0.13 & 2.86$\pm$0.10 & \nodata         & \nodata         & \nodata         & \nodata \nl
            & 9588.61 &       \nodata   & \nodata         & 2.69$\pm$0.10 &  \nodata        & 4.63$\pm$0.11 & \nodata  \nl
            & 9588.62 &       \nodata   & \nodata         &  \nodata        & 3.49$\pm$0.11 &   \nodata       & 4.78$\pm$0.11 \nl
1994 Aug 25 & 9589.53 & 9.45$\pm$0.14 & 6.77$\pm$0.12 & \nodata         & \nodata         & \nodata          & \nodata \nl
            & 9589.54 &       \nodata   & \nodata         & 2.05$\pm$0.10 &  \nodata        & 1.27$\pm$0.10 & \nodata \nl
            & 9589.55 &       \nodata   & \nodata         &   \nodata       & 1.63$\pm$0.10 &  \nodata        & 1.49$\pm$0.10  \nl
1994 Aug 26 & 9590.53 & 8.48$\pm$0.13 & 7.62$\pm$0.13 & \nodata         & \nodata         &  \nodata        & \nodata  \nl
            & 9590.54 &        \nodata  &  \nodata        & 4.44$\pm$0.11 &  \nodata        & 2.87$\pm$0.10 & \nodata  \nl
            & 9590.55 &       \nodata   &   \nodata       &  \nodata        & 3.69$\pm$0.11 &  \nodata        & 2.92$\pm$0.10 \nl
1994 Aug 27 & 9591.35 & 8.44$\pm$0.13 & 8.06$\pm$0.13 & \nodata         & \nodata         & \nodata          &  \nodata\nl
            & 9591.36 &    \nodata      &   \nodata       & 4.78$\pm$0.11 &  \nodata        & 3.66$\pm$0.11 & \nodata \nl
            & 9591.37 &    \nodata      &   \nodata       &     \nodata     & 4.20$\pm$0.11 &        \nodata  & 3.41$\pm$0.13 \nl
1994 Aug 29 & 9593.35 & 7.81$\pm$0.13 & 3.72$\pm$0.11     & 1.97$\pm$0.10 &  \nodata        & 1.40$\pm$0.11 & \nodata \nl
            & 9593.36 &    \nodata      &   \nodata       &    \nodata      & 1.67$\pm$0.10 & \nodata         & 1.53$\pm$0.15 \nl
1994 Sept 1 & 9596.32 & 6.95$\pm$0.12 & 3.35$\pm$0.10 & \nodata         &\nodata          & \nodata          & \nodata \nl
            & 9596.33 &    \nodata      &   \nodata       & 1.76$\pm$0.10 &    \nodata      & 1.63$\pm$0.14 & \nodata \nl
1994 Sept 2 & 9597.61 & 6.46$\pm$0.16 & 3.38$\pm$0.17 & \nodata         & \nodata         & \nodata          & \nodata  \nl
1994 Sept 3 & 9598.07 &    \nodata      &    \nodata      & 2.9$\pm$0.36 &   \nodata       & 0.669$\pm$0.531 &  \nodata \nl
\enddata   

 
\tablenotetext{a}{Expressed as a percentage of the total flux density.}
\tablenotetext{b}{At mid-observation.}

\end{deluxetable}

\clearpage

\begin{deluxetable}{cl@{\hspace{1mm}}c@{\hspace{1mm}}l@{\hspace{8mm}}l@{\hspace{1mm}}c@{\hspace{1mm}}l}
\scriptsize
\tablecaption{Position angle and rotation measure. \label{tbl-parm}}
\tablewidth{0pt}
\tablehead{
\colhead{JD $-$ 2,440,000.5} & \multicolumn{3}{c}{PA~~~~}   & \multicolumn{3}{c}{RM} \\
\colhead{ } & \multicolumn{3}{c}{$\arcdeg$~~~~} & \multicolumn{3}{c}{rad m$^{-2}$}
}
\startdata 
      9579.58  &    41.97 &$\pm$   &1.13\tablenotemark{a}  &7.3 &$\pm$  &6.9\tablenotemark{a} \nl
      9580.50  &    54.5  &$\pm$   &4.6   &     31   &$\pm$  &15 \nl
      9581.64  &    53.48 &$\pm$   &0.99  &     65.3 &$\pm$  &4.1 \nl
      9582.17  &    53.7  &$\pm$   &1.6   &    104.1 &$\pm$  &3.7 \nl
      9582.65  &    49.06 &$\pm$   &0.15  &    59.74 &$\pm$  &0.47 \nl
      9583.15  &    53.20 &$\pm$   &0.27  &    56.70 &$\pm$  &0.65 \nl
      9583.59  &    48.95 &$\pm$   &0.55  &    58.66 &$\pm$  &1.25 \nl
      9584.63  &    44.98 &$\pm$   &0.59  &    61.94 &$\pm$  &0.62 \nl
      9585.64  &    45.65 &$\pm$   &1.48  &    60.18 &$\pm$  &1.20 \nl
      9586.63  &    45.47 &$\pm$   &0.66  &    58.87 &$\pm$  &0.59 \nl
      9587.62  &    42.89 &$\pm$   &0.59  &    58.39 &$\pm$  &0.57 \nl
      9588.62  &    47.53 &$\pm$   &0.96  &    56.38 &$\pm$  &0.79 \nl
      9589.55  &    26.1  &$\pm$   &5.0   &    65.5  &$\pm$  &3.2 \nl
      9590.55  &    38.24 &$\pm$   &0.27  &    63.49 &$\pm$  &0.23 \nl
      9591.37  &    42.92 &$\pm$   &0.32  &    61.80 &$\pm$  &0.29 \nl
      9593.36  &    26.86 &$\pm$   &0.88  &    66.03 &$\pm$  &0.51 \nl
      9596.32  &    19.9  &$\pm$   &3.3\tablenotemark{b}  &67.0 &$\pm$ &2.0\tablenotemark{b} \nl
      9598.07  &    28.0  &$\pm$   &2.8\tablenotemark{b} & 79.9 &$\pm$ &2.1\tablenotemark{b} \nl
\enddata   
\tablenotetext{a}{Average of first four interpolated points.}
\tablenotetext{b}{Obtained from the 1.4, 2.3, 4.8 \& 8.6~GHz points alone.}
\end{deluxetable}

\end{document}